\documentclass[journal]{IEEEtran}
\newtheorem{theorem}{Theorem}{}
{}
\newtheorem{remark}{Remark}{}
\usepackage[]{graphicx}
\usepackage{amsmath}
\usepackage{amssymb}
\usepackage{algorithmic}
\usepackage{algorithm}
\usepackage{caption}
\usepackage{mathtools}
\usepackage{changes}
\usepackage{multirow}
\usepackage{verbatim}
\usepackage{mathrsfs}
\usepackage{bm}
\usepackage{fancyhdr}

\ifCLASSINFOpdf
\else
\fi
\begin{document}
\title{Time-limited pseudo-optimal $\mathcal{H}_2$-model order reduction}
\author{Umair~Zulfiqar,~Victor~Sreeram, and Xin~Du
\thanks{``This paper is a preprint of a paper accepted by IET Control Theory \& Applications and is subject to Institution of Engineering and Technology Copyright.  When the final version is published,
the copy of record will be available at the IET Digital Library".}\thanks{U.~Zulfiqar and V.~Sreeram are with the School of Electrical, Electronics and Computer Engineering, The University of Western Australia, 35 Stirling Highway, Crawley, WA 6009 (email: umair.zulfiqar@research.uwa.edu.au, victor.sreeram@uwa.edu.au).}
\thanks{X.~Du is with the School of Mechatronic Engineering and Automation, Shanghai University, Shanghai 200072, China, and also with the Shanghai Key Laboratory of Power Station Automation Technology, Shanghai University, Shanghai 200444, P.R. China  (e-mail: duxin@shu.edu.cn)}}
\markboth{``\tiny "This paper is a preprint of a paper accepted by IET Control Theory \& Applications and is subject to Institution of Engineering and Technology Copyright.''}%
{Shell \MakeLowercase{\textit{et al.}}:}
\maketitle
\begin{abstract}
A model order reduction algorithm is presented that generates a reduced-order model of the original high-order model, which ensures high-fidelity within the desired time interval. The reduced model satisfies a subset of the first-order optimality conditions for time-limited $\mathcal{H}_2$-model reduction problem. The algorithm uses a computationally efficient Krylov subspace-based framework to generate the reduced model, and it is applicable to large-scale systems. The reduced-order model is parameterized to enforce a subset of the first-order optimality conditions in an iteration-free way. We also propose an adaptive framework of the algorithm, which ensures a monotonic decay in error irrespective of the choice of interpolation points and tangential directions. The efficacy of the algorithm is validated on benchmark model reduction problems.
\end{abstract}
\IEEEpeerreviewmaketitle
\section{Introduction}
\IEEEPARstart{T}{he} intricacies and complexity of the physical systems are increasing each year. The modern-day physical systems are mathematically described by several hundred or thousands of differential equations resulting in a large-scale state-space model. The computing power of the modern-day computers is also increasing at an increasing rate; however, the complexity of the physical systems still poses a computational challenge to the efficient simulation, analysis, and design. Model order reduction (MOR) techniques are used to obtain a reduced-order approximation of the original high-order model, which retains most of its input-output properties. The reduced-order model (ROM) can then be used as a surrogate for the original high-order system in the design and analysis \cite{gugercin2004survey,antoulas2000survey,benner2015survey,reis2008survey,benner2005dimension,obinatamodel}.

Balanced truncation (BT) \cite{moore1981principal} is among the most popular MOR techniques. The preservation of stability, the existence of an a priori error bound, and high accuracy are among the most significant features of BT. In BT, the states which have significant Hankel singular values are retained in the ROM, and the remaining states are truncated. BT requires the solution of two large-scale Lyapunov equations which is a computationally intensive task. The high computational cost of BT hinders its applicability to large-scale systems. Several extensions of BT exist in the literature to reduce its computational cost like \cite{gugercin2003modified,su2002efficient,van2000gramian,li2002low,balakrishnan2001efficient} which suggest replacing the exact solution of the Lyapunov equations with their low-rank approximations. BT is generalized to preserve several other system characteristics like passivity, second-order structure, contractivity, etc. Reference \cite{gugercin2004survey} provides an in-depth survey of BT and its extensions.

The modal configuration is an important mathematical property of the system model, which is related to several physical phenomena. For instance, the interconnected power systems exhibit low-frequency oscillations like local, interplant, and interarea oscillations. These are associated with the modes in the frequency interval of $0-2$ Hz and are poorly damped \cite{kundur1994power}. The small-signal stability analysis and the damping controller design rely heavily on these modes. Various modes can also be associated with the power system components in the network, like generators and power system stabilizers. Therefore, their preservation in the ROM is important from a physical perspective. Moreover, their preservation in the ROM is also beneficial for the accuracy of ROM both in the time and frequency domains \cite{scarciotti2015model,scarciotti2016low,chaniotis2005model,sanchez1996power}. Recently, several eigensolvers are developed which exploit the sparse structure of the large-scale models and efficiently compute the dominant modes \cite{rommes2006efficient,rommes2008computing,rommes2009computing} which are required to be preserved in the ROM for achieving good accuracy. Modal truncation based on these eigensolvers can efficiently generate a ROM for large-scale systems which preserves the dominant modes of the original model. In terms of accuracy, modal truncation is way inferior to BT. However, in many applications, the preservation of important modes of the original system is more important than the overall accuracy in terms of error.

Moment matching is another important class of MOR techniques. In moment matching methods, a ROM is constructed which interpolates the original transfer function and a few of its moments using computationally efficient rational Krylov subspace based-framework. These techniques can easily handle large-scale systems and can be applied even if the model is unknown, and only the input-output data is known \cite{scarciotti2017data}. Moment matching methods have been significantly advanced over the last two decades, and several generalizations and extensions exist which can preserve various system properties while ensuring good accuracy as well. Reference \cite{beattie2014model} provides a detailed survey of moment matching methods.

The $\mathcal{H}_2$-optimal MOR problem is studied extensively in the literature. One popular approach to this problem is to first convert it into an unconstraint minimization problem over Stiefel manifold. Then the ROM is constructed using optimization schemes that minimize the cost function derived over the manifold \cite{yan1999approximate,xu2019unconstrained,wang2018h}. In \cite{yang2019trust}, a trust-region for the $\mathcal{H}_2$-optimal MOR over Stiefel manifold is proposed using the Riemannian trust-region method of \cite{sato2015riemannian}. In \cite{sato2017structure},  the Riemannian trust-region method \cite{sato2015riemannian} is generalized for second-order transfer functions wherein the second-order structure is preserved in the ROM. The second popular approach to the $\mathcal{H}_2$-optimal MOR problem is the tangential interpolation. The first-order optimality conditions (known as Wilson's conditions \cite{wilson1970optimum}) are described as the tangential interpolation conditions in \cite{gugercin2008h_2,van2008h2}. The iterative rational Krylov algorithm (IRKA) \cite{gugercin2008h_2} is considered among the gold standards of these methods, and it efficiently generates a local optimum for the problem. IRKA was first developed for single-input single-output (SISO) systems in \cite{gugercin2008h_2}, and it is generally as accurate as BT even if it is initialized with fairly random interpolation points. It was later generalized for multi-input multi-output (MIMO) systems in \cite{van2008h2}. IRKA is computationally efficient and thus easily applicable to large-scale systems.

In general, the convergence is not guaranteed in IRKA, and it significantly slows down as the number of inputs and outputs increases. Moreover, the stability of the ROM is not guaranteed. There are some methods reported in the literature, which use trust-region methods to speed up the convergence of IRKA like \cite{beattie2009trust,panzer2013greedy,wang2018trust}. In \cite{ibrir2018projection}, a projection-based algorithm is proposed, which provides a sub-optimal solution to the $\mathcal{H}_2$-MOR problem while guaranteeing the stability of the ROM at the same time. The applicability of the algorithm \cite{ibrir2018projection} can also be extended to the nonlinear systems. The Lyapunov equation-based algorithm presented in \cite{wang2010optimal} also guarantees the stability of the ROM while preserving the second-order structure in the ROM. Gugercin presented a modification to IRKA, i.e., iterative SVD rational Krylov algorithm (ISRKA), which satisfies a subset of the first-order optimality conditions for the $\mathcal{H}_2$-MOR problem, and the stability is also guaranteed \cite{gugercin2008iterative}. However, the algorithm presented is an iterative algorithm with no guarantee on convergence. Also, it requires the computation of one large-scale Lyapunov equation which is computationally not feasible in a large-scale setting. In \cite{wolf2014h}, an iteration-free pseudo-optimal rational Krylov (PORK) algorithm is presented, which generates a ROM that satisfies a subset of the first-order optimality conditions like ISRKA, and the ROM is also guaranteed to be stable. Unlike ISKRA, it does not require the solution of a large-scale Lyapunov equation, and thus it is computationally efficient. In \cite{panzer2014model}, a cumulative reduction (CURE) framework is proposed, which constructs the ROM adaptively. In each step, new interpolation conditions are enforced without affecting the conditions induced in the previous steps. An important property of CURE is that if PORK is used in it's every step, the $\mathcal{H}_2$-norm error decays monotonically irrespective of the choice of interpolation points and tangential directions.

Practically, no system or its simulation is run over an infinite time interval. In some applications, a particular time interval is more important in terms of accuracy, e.g., the low-frequency oscillations in the interconnected power system are observed in the first $15$ sec of the simulation, after which these are successfully damped by the power system stabilizers. The first $15$ sec are very important for the small-signal stability analysis of the interconnected power systems \cite{rogers2012power}. Similarly, in a time-limited optimal control problem, the behavior of the system and the performance of the controller is important within a finite time interval \cite{grimble1979solution}. It is, therefore, reasonable to ensure superior accuracy within the desired time interval of the interest. To ensure high-fidelity in a desired limited time interval, BT is generalized to time-limited BT (TLBT) in \cite{gawronski1990model}. TLBT is computationally expensive as it requires the solution of two large-scale dense Lyapunov equations. In \cite{kurschner2018balanced}, the applicability of TLBT is extended to large-scale systems using Krylov subspace-based methods and low-rank approximation of large-scale Lyapunov equations. The ROM is not guaranteed to be stable in TLBT, and an a priori error bound does not exist. In \cite{gugercin2003time}, a modification in TLBT is proposed, which guarantees stability, and an error bound is also defined. However, \cite{gugercin2003time} is computationally more expensive than the original TLBT, and the accuracy is also poor. An $\mathcal{H}_2$-norm error bound for TLBT is also proposed in \cite{redmann2017mathcal}, which is based on the error bound in \cite{redmann2017h2}. TLBT is extended to more general classes of linear systems in \cite{tahavori2013model,haider2017model,haider2019time,zulfiqar2018time} and bilinear systems in \cite{shaker2014time}. In \cite{jazlan2015cross}, a cross Gramian based algorithm for TLBT is presented, which significantly saves the computational time as it requires one instead of two large-scale Lyapunov equations. Further, TLBT is extended to unstable systems in \cite{kumar2017generalized}. In \cite{goyal2019time}, IRKA is generalized to time-limited scenario to approximately achieve the first-order optimality conditions for the time-limited $\mathcal{H}_2$-MOR problem. The first-order optimality conditions for the time-limited $\mathcal{H}_2$-MOR problem are expressed as bi-tangential Hermite interpolation conditions in \cite{sinani2018mathcal}, and a descent-based iterative algorithm is presented for the SISO systems. The algorithm \cite{sinani2018mathcal} generates a ROM that satisfies these conditions for SISO system. It is computationally not feasible in a large-scale setting as it requires the computation of the pole-residue form of the original transfer function, which is an expensive task. Moreover, the stability of the original system is not guaranteed to be preserved in the ROM generated by both the algorithms, and there is no guarantee on the convergence of the algorithms.

In this paper, we present a time-limited MOR algorithm which satisfies a subset of the first-order optimality conditions for the time-limited $\mathcal{H}_2$-MOR problem (as derived in \cite{goyal2019time} and \cite{sinani2018mathcal}). The algorithm is iteration-free, and it guarantees the stability of ROM. The proposed algorithm does not involve any large-scale Lyapunov equation, and it uses a computationally efficient rational Krylov algorithm to construct the ROM. Therefore, it is applicable to large-scale systems. The proposed algorithm uses the parametrization of the ROM approach \cite{astolfi2010model,ahmad2011krylov} to enforce a subset of the optimality conditions and to place the poles and their associated input or output residues to the specified locations. Thus, it can also preserve the modes and their associated input or output residues of the original systems like modal truncation. Therefore, the proposed algorithm can also be used for the applications wherein modal preservation is an important property to be preserved in the ROM. We also present an adaptive version of the algorithm, which ensures that the time-limited $\mathcal{H}_2$-norm error decays monotonically irrespective of the choice of interpolation points and tangential directions. The proposed algorithm also provides approximate time-limited Gramians, which monotonically approach to the exact solutions. The approximate Gramians can be used for a computationally efficient implementation of TLBT. We have tested our algorithm on benchmark MOR problems, and the simulation results confirm the efficacy of the proposed algorithm.
\section{Preliminaries}
Let $H(s)$ be the $n^{th}$ order model of the original high-order system. The MOR problem is to find a $r^{th}$ ($r<<n$) order ROM $\hat{H}_r(s)$ of $H(s)$ such that the error $||H(s)-\hat{H}_r(s)||$ is small in some defined sense. Let $x\in\mathbb{R}^{n\times 1}$, $u\in\mathbb{R}^{1\times m}$, and $y\in\mathbb{R}^{p\times 1}$ be the state, input, and output vectors of the state-space representation of $H(s)$, i.e.,
\begin{align}
\dot{x}&=Ax+Bu,\hspace*{1cm}y=Cx,\nonumber\\
H(s)&=C(sI-A)^{-1}B\label{EQ1}.
\end{align}
Let $x_r\in\mathbb{R}^{r\times 1}$ and $y_r\in\mathbb{R}^{p\times 1}$ be the state and output vectors of the state-space representation of $\hat{H}_r(s)$, i.e.,
\begin{align}
\dot{x}_r&=\hat{A}_rx_r+\hat{B}_ru\hspace*{1cm}\hat{y}_r=\hat{C}_rx_r.\label{EQ2}
\end{align}
$\hat{W}_r$ and $\hat{V}_r$ project $H(s)$ onto a reduced subspace such that the dominant dynamics of $H(s)$ are retained in $\hat{H}_r(s)$, i.e.,
\begin{align}
	\hat{H}_r(s)&=C\hat{V}_r(sI-\hat{W}_r^TA\hat{V}_r)^{-1}\hat{W}_r^TB\nonumber\\
	&=\hat{C}_r(sI-\hat{A}_r)^{-1}\hat{B}_r.\nonumber
\end{align}
The time-limited MOR problem is to compute a ROM $(\hat{A}_r,\hat{B}_r,\hat{C}_r)$ such that $y\approx \hat{y}_r$ within the desired time interval $[0,t]$ sec. The important mathematical notations which are used throughout the text are tabulated in Table \ref{tab1}.
\begin{table}[!h]
	\centering
	\caption{Mathematical Notations}\label{tab1}
	\begin{tabular}{|c|c|}
		\hline
		Notation & Meaning \\ \hline
		$\begin{bmatrix}.\end{bmatrix}^*$   & Hermitian of the matrix\\
        $tr(\cdot)$& Trace of the matrix\\
		$\lambda_i(\cdot)$&Eigenvalues of the matrix\\
		$Ran(\cdot)$&Range of the matrix\\
		$orth(\cdot)$&Orthogonal basis for the range of the matrix\\
		$\underset {i=1,\cdots,r}{span}\{\cdot\}$&Span of the set of $r$ vectors\\\hline
	\end{tabular}
\end{table}
\\Let $h(t)=Ce^{At}B$ and $\hat{h}_r(t)=\hat{C}_re^{\hat{A}_rt}\hat{B}_r$ be the impulse responses of $H(s)$ and $\hat{H}_r(s)$, respectively. The fourier transforms of $h(t)$ and $\hat{h}_r(t)$ are given by $CF(\omega)B$ and $\hat{C}_r\hat{F}(\omega)\hat{B}_r$, respectively where $F(\omega)=(j\omega I-A)^{-1}$ and $\hat{F}(\omega)=(j\omega I-\hat{A}_r)^{-1}$. The controllability Gramian $P$ and observability Gramian $Q$ of the state-space realization $(A,B,C)$ are defined as the following
\begin{align}
P&=\int_{0}^{\infty}e^{A\tau}BB^Te^{A^T\tau}d\tau=\frac{1}{2\pi}\int_{-\infty}^{\infty}F(\nu)BB^TF^*(\nu)d\nu\\
Q&=\int_{0}^{\infty}e^{A^T\tau}C^TCe^{A\tau}d\tau=\frac{1}{2\pi}\int_{-\infty}^{\infty}F^*(\nu)C^TCF(\nu)d\nu.
\end{align}
The time-limited controllability Gramian $P_T$ and the time-limited observability Gramain $Q_T$ defined over the time interval $[0,t]$ \cite{gawronski1990model} are defined as
\begin{align}
	P_T&=\int_{0}^{t}e^{A\tau}BB^Te^{A^T\tau}d\tau\nonumber\\
    Q_T&=\int_{0}^{t}e^{A^T\tau}C^TCe^{A\tau}d\tau.\nonumber
\end{align}
Similarly, the frequency-limited controllability Gramian $P_\Omega$ and the frequency-limited observability Gramain $Q_\Omega$ defined over the frequency interval $[0,\omega]$ \cite{gawronski1990model} are defined as
\begin{align}
P_\Omega&=\frac{1}{2\pi}\int_{-\omega}^{\omega}F(\nu)BB^TF^*(\nu)d\nu\\
Q_\Omega&=\frac{1}{2\pi}\int_{-\omega}^{\omega}F^*(\nu)C^TCF(\nu)d\nu.
\end{align}
\textbf{Definition 1.} The $\mathcal{H}_2$-norm \cite{gugercin2008h_2} of $H(s)$ is defined as
\begin{align}
||H(s)||_{\mathcal{H}_2}&=\sqrt{tr\Big(\int_{0}^{\infty}h(\tau)h^T(\tau)d\tau\Big)}\nonumber\\
&=\sqrt{tr\Big(\int_{0}^{\infty}Ce^{A\tau}BB^Te^{A^T\tau}C^Td\tau\Big)\nonumber}\\
&=\sqrt{tr\big(CPC^T\big)}\nonumber\\
&=\sqrt{tr\Big(\int_{0}^{\infty}h^T(\tau)h(\tau)d\tau\Big)}\nonumber\\
&=\sqrt{tr\Big(\int_{0}^{\infty}B^Te^{A^T\tau}C^TCe^{A\tau}Bd\tau\Big)\nonumber}\\
&=\sqrt{tr\big(B^TQB\big)}.\nonumber
\end{align}
\textbf{Definition 2.} The time-limited $\mathcal{H}_2$-norm \cite{goyal2019time} for the time interval $[0,t]$, i.e., $\mathcal{H}_{2,t}$ of $H(s)$ is defined as
\begin{align}
||H(s)||_{\mathcal{H}_{2,t}}&=\sqrt{tr\Big(\int_{0}^{t}h(\tau)h^T(\tau)d\tau\Big)}\nonumber\\
&=\sqrt{tr\Big(\int_{0}^{t}Ce^{A\tau}BB^Te^{A^T\tau}C^Td\tau\Big)}\nonumber\\
&=\sqrt{tr\big(CP_TC^T\big)}\nonumber\\
&=\sqrt{tr\Big(\int_{0}^{t}h^T(\tau)h(\tau)d\tau\Big)}\nonumber\\
&=\sqrt{tr\Big(\int_{0}^{t}B^Te^{A^T\tau}C^TCe^{A\tau}Bd\tau\Big)}\nonumber\\
&=\sqrt{tr\big(B^TQ_TB\big)}.\nonumber
\end{align}
\textbf{Definition 3.} The frequency-limited $\mathcal{H}_2$-norm \cite{petersson2014model} for the frequency interval $[0,\omega]$, i.e., $\mathcal{H}_{2,\omega}$ of $H(s)$ is defined as
\begin{align}
||H(s)||_{\mathcal{H}_{2,\omega}}&=\sqrt{tr\Big(\frac{1}{2\pi}\int_{-\omega}^{\omega}H(j\nu)H^*(j\nu)d\nu\Big)}\nonumber\\
&=\sqrt{tr\Big(\frac{1}{2\pi}\int_{-\omega}^{\omega}CF(\nu)BB^TF^*(\nu)C^Td\nu\Big)}\nonumber\\
&=\sqrt{tr\big(CP_\Omega C^T\big)}\nonumber\\
&=\sqrt{tr\Big(\frac{1}{2\pi}\int_{-\omega}^{\omega}H^*(j\nu)H(j\nu)d\nu\Big)}\nonumber\\
&=\sqrt{tr\Big(\frac{1}{2\pi}\int_{-\omega}^{\omega}B^TF^*(\nu)C^TCF(\nu)Bd\nu\Big)}\nonumber\\
&=\sqrt{tr\big(B^TQ_\Omega B\big)}.\nonumber
\end{align}
\textbf{Definition 4.} The $\mathcal{H}_{2,t}$-norm \cite{goyal2019time} of the error system $E(s)=H(s)-\hat{H}_r(s)$ is given by
\begin{align}
	||&E(s)||_{\mathcal{H}_{2,t}}\nonumber\\
	&=\sqrt{tr(CP_TC^T)-2tr(C\tilde{P}_T \hat{C}_r^T)+tr(\hat{C}_r\hat{P}_T\hat{C}_r^T)}\nonumber\\
	&=\sqrt{tr(B^TQ_TB)-2tr(\hat{B}_r^T\tilde{Q}_T B)+tr(\hat{B}_r^T\hat{Q}_T\hat{B}_r)}\nonumber
\end{align} where
\begin{align}
AP_T+P_T A^T+BB^T-e^{At}BB^Te^{A^Tt}&=0,\label{EQ7}\\
	A^TQ_T+Q_T A+C^TC-e^{A^Tt}C^TCe^{At}&=0,\\ \hat{A}_r\hat{P}_T+\hat{P}_T\hat{A}_r^T+\hat{B}_r\hat{B}_r^T-e^{\hat{A}_rt}\hat{B}_r\hat{B}_r^Te^{\hat{A}_r^Tt}&=0,\label{eq:8b}\\
	\hat{A}_r^T\hat{Q}_T+\hat{Q}_T\hat{A}_r+\hat{C}_r^T\hat{C}_r-e^{\hat{A}_r^Tt}\hat{C}_r^T\hat{C}_re^{\hat{A}_rt}&=0,\label{eq:9b}\\
	A\tilde{P}_T+\tilde{P}_T\hat{A}_r^T+B\hat{B}_r^T-e^{At}B\hat{B}_r^Te^{\hat{A}_r^Tt}&=0,\\
	\hat{A}_r^T\tilde{Q}_T+\tilde{Q}_TA+\hat{C}_r^TC-e^{\hat{A}_r^Tt}\hat{C}_r^TCe^{At}&=0.\label{EQ12}
\end{align} See \cite{goyal2019time} for the proof.
\subsection{PORK \cite{wolf2014h}}
$\hat{H}_r(s)$ interpolates $H(s)$ at the interpolation points $\sigma_i$ in the respective (right) tangential directions $\hat{c}_i\in\mathbb{C}^{m\times1}$ for any output rational Krylov subspace $\hat{W}_r$ such that $\hat{W}_r^T\hat{V}_r=I$ if the input rational Krylov subspace $\hat{V}_r$ satisfies the following property
\begin{align}
Ran(\hat{V}_r)=\underset {i=1,\cdots,r}{span}\{(\sigma_iI-A)^{-1}B \hat{c}_i\}.
\end{align}
Choose $\hat{W}_r$ as $\hat{W}_r=\hat{V}_r(\hat{V}_r\hat{V}_r^T)^{-1}$, and compute the following matrices
	\begin{align}
\hat{A}_r&=\hat{W}_r^TA\hat{V}_r,\hspace*{0.3cm}\hat{B}_r=\hat{W}_r^TB,\\
	B_\bot&=B-\hat{V}_r\hat{B}_r,\\
	\hat{L}_r&=(B_\bot^TB_\bot)^{-1}B_\bot^T\big(A\hat{V}_r-\hat{V}_r\hat{A}_r\big),\\
	\hat{S}_r&=\hat{A}_r-\hat{B}_r \hat{L}_r.
	\label{eq:7b}\end{align}
Then $\hat{V}_r$ satisfies the following Sylvester equations:
\begin{align}
A\hat{V}_r+\hat{V}_r(-\hat{S}_r)+B(-\hat{L}_r)=0,\label{EQ23}\\
A\hat{V}_r+\hat{V}_r(-\hat{A}_r)+B_{\perp}(-\hat{L}_r)=0\label{EQ24}
\end{align} where $\{\sigma_1,\cdots,\sigma_r\}$ are the eigenvalues of $\hat{S}_r$. A family of ROMs which satisfy the interpolation condition
\begin{align}\hat{H}_r(\sigma_i)\hat{c}_i=H(\sigma_i)\hat{c}_i\label{EQ25}
\end{align} can be obtained by parametrizing the ROM in $\hat{B}_r$ if all the interpolation points $\mathbf{\sigma}_i$ have positive real parts, and $(\hat{S}_r,\hat{L}_r)$ is observable, i.e.,
\begin{align}
	\hat{A}_r=\hat{S}_r+\xi\hat{L}_r,&& \hat{B}_r=\xi,&& \hat{C}_r=C\hat{V}_r.\nonumber
\end{align}
$\hat{H}_r(s)$ ensures that \begin{align}||H(s)-\hat{H}_r(s)||^2_{\mathcal{H}_\mathrm{2}}=||H(s)||^2_{\mathcal{H}_\mathrm{2}}-||\hat{H}_r(s)||^2_{\mathcal{H}_\mathrm{2}}\label{eq:4}
\end{align}
if $\xi$ is set to $\xi=-\hat{Q}_s^{-1}\hat{L}_r^T$ where $\hat{Q}_s$ solves
\begin{align}
	-\hat{S}_r^T\hat{Q}_s-\hat{Q}_s\hat{S}_r+\hat{L}_r^T\hat{L}_r=0.\nonumber
\end{align}
The equation (\ref{eq:4}) is a subset of the first-order optimality conditions for the local optimality problem $||H(s)-\hat{H}_r(s)||^2_{\mathcal{H}_\mathrm{2}}$, and thus $\hat{H}_r(s)$ is a pseudo-optimal ROM of $H(s)$.
\subsection{CURE \cite{panzer2014model}}
The $r^{th}$-order ROM in CURE is constructed adaptively in $k$ steps. $(\hat{A}_r,\hat{B}_r,\hat{C}_r)$, $\hat{S}_r$, $\hat{L}_r$, and $B_{\perp}$ are accumulated after each step such that the interpolation condition (\ref{EQ25}), and the Sylvester equations (\ref{EQ23}) and (\ref{EQ24}) are satisfied after each step. In other words, at each step, new interpolation points and the tangential directions are added without disturbing the interpolation conditions induced by the previous interpolation points and tangential directions.\\
Let $\hat{V}^{(k)}$, $\hat{S}^{(k)}$, $\hat{L}^{(k)}$, $(\hat{A}^{(k)},\hat{B}^{(k)},\hat{C}^{(k)})$, and $B_{\perp}^{(k)}$ be matrices of $k^{th}$ step for any $\hat{W}^{(k)}$ such that $(\hat{W}^{(k)})^T\hat{V}^{(k)}=I$, i.e.,
\begin{align}
\hat{A}^{(k)}&=(\hat{W}^{(k)})^TA\hat{V}^{(k)},&& \hat{B}^{(k)}=(\hat{W}^{(k)})^TB_{\perp}^{(k-1)},\nonumber\\
\hat{C}^{(k)}&=C\hat{V}^{(k)}\nonumber
\end{align} where
\begin{align}
A\hat{V}^{(k)}-\hat{V}^{(k)}\hat{S}^{(k)}-B_{\perp}^{(k-1)}\hat{L}^{(k)}=0,\\
A\hat{V}^{(k)}-\hat{V}^{(k)}\hat{A}^{(k)}-B_{\perp}^{(k)}\hat{L}^{(k)}=0,
\end{align} $B_{\perp}^{(0)}=B$, and $B_{\perp}^{(k)}=B_{\perp}^{(k-1)}-\hat{V}^{(k)}\hat{B}^{(k)}$.\\
After each step, the ROM can be accumulated, and a $r^{th}$-order ROM can be obtained adaptively in $k$ steps, i.e., the order of the ROM grows after each step without affecting the interpolation conditions induced in the previous steps. The accumulated ROM and the associated matrices for $i=1,\cdots,k$ are given by
\begin{align}
A_{tot}^{(i)}&=\begin{bmatrix}A_{tot}^{(i-1)}&0\\&\\\hat{B}^{(i)}L_{tot}^{(i-1)}&\hat{A}^{(i)}\end{bmatrix},&& B_{tot}^{(i)}=\begin{bmatrix}B_{tot}^{(i-1)}\\\\\hat{B}^{(i)}\end{bmatrix},\nonumber\\
C_{tot}^{(i)}&=\begin{bmatrix}C_{tot}^{(i-1)}&&\hat{C}^{(i)}\end{bmatrix},&&L_{tot}^{(i)}=\begin{bmatrix}L_{tot}^{(i-1)}&&\hat{L}^{(i)}\end{bmatrix},\nonumber\\
S_{tot}^{(i)}&=\begin{bmatrix}S_{tot}^{(i-1)}&-B_{tot}^{(i-1)}\hat{L}^{(i)}\\&\\0&\hat{S}^{(i)}\end{bmatrix},&&V_{tot}^{(i)}=\begin{bmatrix}V_{tot}^{(i-1)}& \hat{V}^{(i)}\end{bmatrix}\nonumber
\end{align} where
\begin{align}
AV_{tot}^{(i)}-V_{tot}^{(i)}S_{tot}^{(i)}-BL_{tot}^{(i)}=0,\nonumber
\end{align} and $A_{tot}^{(0)}$, $B_{tot}^{(0)}$, $C_{tot}^{(0)}$, $L_{tot}^{(0)}$, $S_{tot}^{(0)}$, and $V_{tot}^{(0)}$ are all empty matrices.\\
An important property of CURE is that if $(\hat{A}^{(i)},\hat{B}^{(i)},\hat{C}^{(i)})$ is computed using PORK for $i=1,\cdots,k$, $(A_{tot}^{(i)},B_{tot}^{(i)},C_{tot}^{(i)})$ stays pseudo-optimal. This further implies that $(A_{tot}^{(i-1)},B_{tot}^{(i-1)},C_{tot}^{(i-1)})$ is a pseudo-optimal ROM of $(A_{tot}^{(i)},B_{tot}^{(i)},C_{tot}^{(i)})$ as $S_{tot}^{(i)}$ and $\hat{C}_{tot}^{(i)}$ contain the interpolation points and tangential directions encoded in $S_{tot}^{(i-1)}$ and $\hat{C}_{tot}^{(i-1)}$, respectively. Let $H_{tot}^{(i)}\big(s\big)$ be defined as
\begin{align}
H_{tot}^{(i)}\big(s\big)=C_{tot}^{(i)}(sI-A_{tot}^{(i)})^{-1}B_{tot}^{(i)}.\nonumber
\end{align}
If $H_{tot}^{(i)}\big(s\big)$ stays pesudo-optimal, the following holds
\begin{align}
||H(s)-H_{tot}^{(i)}\big(s\big)||^2_{\mathcal{H}_2}=||H(s)||^2_{\mathcal{H}_2}-||H_{tot}^{(i)}\big(s\big)||^2_{\mathcal{H}_2}.\nonumber
\end{align}
Thus $||H(s)||^2_{\mathcal{H}_2}\geq||H_{tot}^{(i)}\big(s\big)||^2_{\mathcal{H}_2}$. Moreover, since
\begin{align}
||H_{tot}^{(i)}\big(s\big)-H_{tot}^{(i-1)}\big(s\big)||^2_{\mathcal{H}_2}=||H_{tot}^{(i)}\big(s\big)||^2_{\mathcal{H}_2}-||H_{tot}^{(i-1)}\big(s\big)||^2_{\mathcal{H}_2},\nonumber
\end{align} $||H_{tot}^{(i)}\big(s\big)||^2_{\mathcal{H}_2}\geq ||H_{tot}^{(i-1)}\big(s\big)||^2_{\mathcal{H}_2}$. Therefore, $||H(s)-H_{tot}^{(i)}\big(s\big)||^2_{\mathcal{H}_2}$ decays monotonically irrespective of the choice of interpolation points and tangential directions.
\subsection{TLBT \cite{gawronski1990model}}
TLBT \cite{gawronski1990model} is a generalization of BT \cite{moore1981principal} wherein the standard controllability and observability Gramians, which are defined over the infinite time horizon, are replaced with the ones defined over the time interval of interest. $\hat{V}_r$ and $\hat{W}_r$ in TLBT are computed as $\hat{V}_r=\hat{T}_r\hat{R}_r^T$ and $\hat{W}_r=\hat{T}_{r}^{-T}\hat{R}_r^T$, respectively,
where $\hat{R}_r=\begin{bmatrix}I_{r\times r} & 0_{r\times (n-r)}\nonumber\end{bmatrix}$, $\hat{T}_r^{-1}P_T \hat{T}_r^{-T}=\hat{T}_r^TQ_T \hat{T}_r=diag( \hat{\sigma}_1, \hat{\sigma}_2,\cdots,\hat{\sigma}_n)$, and $\hat{\sigma}_1 \geq \hat{\sigma}_2 \geq \cdots \geq \hat{\sigma}_n$. The ROM is then obtained as
\begin{align}
\hat{A}_r=\hat{W}_r^TA\hat{V}_r,&&\hat{B}_r=\hat{W}_r^TB,&&\hat{C}_r=C\hat{V}_r.\nonumber
\end{align}
\subsection{$\mathcal{H}_{2,t}$-Optimal MOR} Define $G(s)$ and $\hat{G}_r(s)$ as the following
\begin{align}
G(s)&=-e^{-st_f}C(sI-A)^{-1}e^{At_f}B+H(s),\\
\hat{G}_r(s)&=-e^{-st_f}\hat{C}_r(sI-\hat{A}_r)^{-1}e^{\hat{A}_rt_f}\hat{B}_r+\hat{H}_r(s).\label{EQ30}
\end{align}
Let $\hat{H}_r(s)$ be represented in the following pole-residue form
\begin{align}
\hat{H}_r(s)=\sum_{k=1}^{r}\frac{\hat{l}_k\hat{r}_k^T}{s-\hat{\lambda}_k}.
\end{align}
It is shown in \cite{sinani2018mathcal} that $\hat{H}_r(s)$ is a local optimum for the problem $||H(s)-\hat{H}_r(s)||^2_{\mathcal{H}_{2,t}}$ if the following bi-tangential Hermite interpolation conditions are satisfied
\begin{align}
\hat{l}_k^TG(-\hat{\lambda}_k)&=\hat{l}_k^T\hat{G}_r(-\hat{\lambda}_k)\label{EQ32}\\
G(-\hat{\lambda}_k)\hat{r}_k&=\hat{G}_r(-\hat{\lambda}_k)\hat{r}_k\label{EQ33}\\
\hat{l}_k^TG^\prime(-\hat{\lambda}_k)\hat{r}_k&=\hat{l}_k^T\hat{G}_r^{\prime}(-\hat{\lambda}_k)\hat{r}_k.\label{EQ34}
\end{align}
Further, it is shown in \cite{goyal2019time} that these are equivalent to the following Gramian-based first-order optimality conditions
\begin{align}
	\hat{C}_r\hat{P}_T&=C\tilde{P}_T\label{eq:9}\\
	\hat{Q}_T\hat{B}_r&=\tilde{Q}_TB\label{eq:10}\\
	\hat{e}_i^*\tilde{X}_{\infty}[\tilde{P}_T-te^{At}B\hat{B}_r^*e^{\hat{A}_r^*t}]\hat{e}_i&=\nonumber\\ \hat{e}_i^*\hat{Q}_{\infty}[\hat{P}_T&-te^{\hat{A}_rt}\hat{B}_r\hat{B}_r^*e^{\hat{A}_r^*t}]\hat{e}_i\label{11c}
\end{align} where
\begin{align}
	\hat{A}_r\tilde{X}_{\infty}+\tilde{X}_{\infty}A^*+\hat{C}_r^*C=0,\nonumber\\
	\hat{A}_r^*\hat{Q}_{\infty}+\hat{Q}_{\infty}\hat{A}_r+\hat{C}_r^*\hat{C}_r=0,\nonumber
\end{align} and $\hat{e}_i$ is the unit vector.
When either the equation (\ref{eq:9}) or (\ref{eq:10}) (or equivalently the equation (\ref{EQ32}) or (\ref{EQ33})) is satisfied, the following holds
\begin{align} ||H(s)-\hat{H}_r(s)||^2_{\mathcal{H}_{2,t}}=||H(s)||^2_{\mathcal{H}_{2,t}}-||\hat{H}_r(s)||^2_{\mathcal{H}_{2,t}}\label{eq:13}.
\end{align}
\subsection{Issues in $\mathcal{H}_{2,t}$-Optimal MOR}
$\hat{H}_r(s)$ and resultantly $\{\hat{\lambda}_k,\hat{l}_k,\hat{r}_k\}_{k=1,\cdots,r}$ are not known a priori which necessitate an iterative algorithm starting with a random guess of the interpolation data. The bi-tangential Hermite interpolation conditions (\ref{EQ32})-(\ref{EQ34}) cannot be achieved if IRKA is applied to $G(s)$. Note that a ROM of $G(s)$ is not required. A ROM $\hat{H}_r(s)$ is required which satisfies (\ref{EQ32})-(\ref{EQ34}). If IRKA is applied to $G(s)$, there is no way to extract $\hat{H}_r(s)$ from $\hat{G}_r(s)$ as IRKA cannot preserve the structure of $\hat{G}_r(s)$ according to (\ref{EQ30}). In \cite{sinani2018mathcal}, a decent algorithm is presented for SISO systems which computes $\hat{H}_r(s)$ which satisfies $G(-\hat{\lambda}_k)=\hat{G}_r(-\hat{\lambda}_k)$ and $G^\prime(-\hat{\lambda}_k)=\hat{G}_r^{\prime}(-\hat{\lambda}_k)$ using quasi-Newton type optimization. This is a computationally expensive strategy to achieve the optimality conditions, which is just limited to the SISO systems.\\
In \cite{goyal2019time}, a projection-based algorithm the time-limited IRKA (TLIRKA) is presented, which tends to achieve optimality conditions (\ref{eq:9})-(\ref{11c}). The algorithm does not achieve an exact local optimum; nevertheless, it yields a ROM, which exhibits high-fidelity within the specified time interval $[0,t]$. Let $(\hat{A}_r,\hat{B}_r,\hat{C}_r)$ be the initial guess of the ROM, and let $QSQ^{-1}$ be the spectral factorization of $\hat{A}_r$, $\mathcal{B}=S\hat{B}_r$, and $\mathcal{C}=\hat{C}_rS^{-1}$ where $Q=\begin{bmatrix}q_1&\cdots&q_r\end{bmatrix}$ and $q_i$ is the eigenvector of $\lambda_i(\hat{A}_r)$. The algorithm in \cite{goyal2019time} heuristically suggests replacing $B\mathcal{B}^*$ and $C^*\mathcal{C}$ in IRKA with $B\mathcal{B}^*-e^{At}B\mathcal{B}^*e^{St}$ and $C^*\mathcal{C}-e^{A^*t}C^*\mathcal{C}e^{St}$, respectively for the calculation of the reduction subspaces (see the equations (\ref{EQ40})-(\ref{EQ41})). The input and output subspaces in TLIRKA \cite{goyal2019time} are computed as
\begin{align}
A\hat{V}_{r,t}+\hat{V}_{r,t}S+B\mathcal{B}^*-e^{At}B\mathcal{B}^*e^{St}=0,\label{EQ40}\\
A^*\hat{W}_{r,t}+\hat{W}_{r,t}S+C^*\mathcal{C}-e^{A^*t}C^*\mathcal{C}e^{St}=0.\label{EQ41}
\end{align}
$\hat{V}_{r,t}$ and $\hat{W}_{r,t}$ is set to $orth(\hat{V}_{r,t})$ and $orth(\hat{W}_{r,t})$, respectively, and the ROM is updated as
\begin{align}
\hat{A}_r&=(\hat{W}_{r,t}^*\hat{V}_{r,t})^{-1}\hat{W}_{r,t}^*A\hat{V}_{r,t}, &&\hat{B}_r=(\hat{W}_{r,t}^*\hat{V}_{r,t})^{-1}\hat{W}_{r,t}^*B,\nonumber\\
\hat{C}_r&=C\hat{V}_{r,t}.\nonumber
\end{align} The process is repeated until the relative change in $S$ stagnates.\\
We will explain in the next section why this simple heuristic modification in IRKA generates a high-fidelity ROM despite the fact that it does not achieve the optimality conditions (\ref{eq:9})-(\ref{11c}). Here we just mention that TLIRKA tends to achieve (\ref{eq:9}) and (\ref{eq:10}) upon convergence; however, it fails to do so because a particular structure is required to be preserved in the ROM, which is not possible in an IRKA-type framework. Moreover, the convergence is not guaranteed in both the aforementioned approaches.
\section{Main Results}
In this section, we present an iteration-free rational Krylov subspace based MOR algorithm, which generates a time-limited pseudo-optimal ROM of $H(s)$. We call a ROM as time-limited pseudo-optimal if it satisfies (\ref{eq:13}). The ROM thus satisfies a subset of the optimality conditions for the problem $||H(s)-\hat{H}_r(s)||^2_{\mathcal{H}_{2,t}}$, i.e., $\hat{H}_r(s)$ satisfies either (\ref{eq:9}) or (\ref{eq:10}). We name our algorithm as ``time-limited PORK (TLPORK)''. We then generalize CURE such that $||H(s)-\hat{H}_r(s)||^2_{\mathcal{H}_{2,t}}$ decays monotonically after each step irrespective of the choice of interpolation points and tangential directions, and we name it as ``time-limited CURE (TLCURE)''.
\subsection{TLPORK}
Let us define $H_T(s)$, $G_T(s)$, $\hat{H}_{T_r}(s)$, and $\hat{G}_{T_r}(s)$ as the following
\begin{align}
H_T(s)&=C(sI-A)^{-1}B_T,\nonumber\\
G_T(s)&=C_T(sI-A)^{-1}B,\nonumber\\
\hat{H}_{T_r}(s)&=\hat{C}_r(sI-\hat{A}_r)^{-1}\hat{B}_{T_r},\nonumber\\
\hat{G}_{T_r}(s)&=\hat{C}_{T_r}(sI-\hat{A}_r)^{-1}\hat{B}_{r}\nonumber
\end{align} where
\begin{align}
B_T&=\begin{bmatrix}B & -e^{At}B\end{bmatrix},\hspace*{1cm}\hat{B}_{T_r}=\begin{bmatrix}\hat{B}_r & -e^{\hat{A}_rt}\hat{B}_r\end{bmatrix},\label{EQ42}\\
C_T&=\begin{bmatrix}C \\ -Ce^{At}\end{bmatrix},\textnormal{ and }\hspace*{0.88cm}\hat{C}_{T_r}=\begin{bmatrix}\hat{C}_r \\ -\hat{C}_re^{\hat{A}_rt}\end{bmatrix}.\label{EQ43}
\end{align}
We begin our investigation by finding out the reason why TLIRKA constructs a nearly-optimal ROM due to the heuristic choice of the reduction subspaces proposed in \cite{goyal2019time}. Intuitively, it appears that TLIRKA tries to satisfy the following interpolation conditions
\begin{align}
H_T(\hat{\lambda}_k)r_k&=\hat{H}_{T_r}(-\hat{\lambda}_k)r_k,\label{EQ44}\\
l_k^TG_T(-\hat{\lambda}_k)&=l_k^T\hat{G}_{T_r}(-\hat{\lambda}_k)\label{EQ45}
\end{align}
while maintaining the structure of $\hat{B}_{T_r}$ and $\hat{C}_{T_r}$ where
\begin{align}
\begin{bmatrix}\begin{matrix}\hat{r}_1&\cdots&\hat{r}_r\end{matrix}\\\begin{bmatrix}\hat{r}_1&\cdots&\hat{r}_r\end{bmatrix}e^{diag(\hat{\lambda}_1,\cdots,\hat{\lambda}_r)t}\end{bmatrix}&=\begin{bmatrix}r_1&\dots&r_r\end{bmatrix},\nonumber\\
\begin{bmatrix}\begin{matrix}\hat{l}_1\\\vdots\\\hat{l}_r\end{matrix}&e^{diag(\hat{\lambda}_1,\cdots,\hat{\lambda}_r)t}\begin{bmatrix}\hat{l}_1\\\vdots\\\hat{l}_r\end{bmatrix}\end{bmatrix}&=\begin{bmatrix}l_1\\\vdots\\l_r\end{bmatrix}.\nonumber\end{align} However, it does not retain that structure because this is not possible in an IRKA-type framework. Our first aim is to construct a ROM, which satisfies (\ref{EQ44}) or (\ref{EQ45}) while maintaining the structure of $\hat{B}_{T_r}$ or $\hat{C}_{T_r}$ according to (\ref{EQ42}) or (\ref{EQ43}), respectively. The next aim is to study the properties of such ROM and investigate its connection with the optimality conditions (\ref{eq:9})-(\ref{11c}).\\
Let us define $\hat{S}_r$, $\hat{L}_r$, and $L_T$ as
\begin{align}
\hat{S}_r&=diag(\sigma_1,\cdots,\sigma_r),\hspace*{0.4cm}\hat{L}_r=\begin{bmatrix}\hat{c}_1&\cdots&\hat{c}_r\end{bmatrix},\nonumber\\ L_T&=\begin{bmatrix}\hat{L}_r\\\hat{L}_re^{-\hat{S}t}\end{bmatrix}=\begin{bmatrix}c_1&\dots&c_r\end{bmatrix}.\label{eq:14}
\end{align}
$\hat{H}_{T,r}(s)$ interpolates $H_T(s)$ at the interpolation points $\sigma_k$ in the respective (right) tangential directions $c_k\in\mathbb{C}^{2m\times1}$ for any output rational Krylov subspace $\hat{W}_{r,t}$ such that $\hat{W}_{r,t}^*\hat{V}_{r,t}=I$ if the input rational Krylov subspace $\hat{V}_{r,t}$ is defined as
\begin{align} \hat{V}_{r,t}=\begin{bmatrix}(A-\sigma_1I)^{-1}B_Tc_1&\cdots&(A-\sigma_rI)^{-1}B_Tc_r\end{bmatrix}\nonumber.
\end{align}
Owing to the relation with the Sylvester equation \cite{panzer2014model}, $\hat{V}_{r,t}$ solves the following Sylvester equation
\begin{align}
	A\hat{V}_{r,t}+\hat{V}_{r,t}(-\hat{S}_r)+B_T(-L_T)=0.\label{eq:15}
\end{align}
A family of ROMs, which satisfy the interpolation condition $\hat{H}_{T_r}(\sigma_k)c_k=H_T(\sigma_k)c_k$, can be obtained by parametrizing $\hat{H}_{T_r}(\sigma_k)$ in $\hat{B}_T$ if all the interpolation points $\sigma_k$ have positive real parts, and $(\hat{S}_r,L_T)$ is observable, i.e.,
\begin{align}
	\hat{A}_r=\hat{S}_r+\xi L_T&& \hat{B}_T=\xi&& \hat{C}_r=C\hat{V}_{r,t}.\nonumber
\end{align}
This can be verified by multiplying (\ref{eq:15}) with $\hat{W}_{r,t}^*$ from the left; see \cite{panzer2014model} for more details. The assumption on the interpolation points to have positive real parts is to ensure that the ROM is stable as we aim to place the poles at their mirror images, and the assumption on the observability of the pair $(\hat{S}_r,L_T)$ is to ensure that the pole-placement is possible. We now propose a novel choice of $\xi$ which retains the structure of $\hat{B}_{T_r}$ according to the equation (\ref{EQ42}), i.e.,
\begin{align}
	\xi=\begin{bmatrix}-\hat{Q}_S^{-1}\hat{L}_r^*&\hat{Q}_S^{-1}e^{-\hat{S}_r^*t}\hat{L}_r^*\end{bmatrix}
\end{align} where
\begin{align}
	-\hat{S}_r^*\hat{Q}_S-\hat{Q}_S\hat{S}_r+\hat{L}_r^*\hat{L}_r-e^{-\hat{S}_r^*t}\hat{L}_r^*\hat{L}_re^{-\hat{S}_rt}=0.\label{eq:17}
\end{align}
$\hat{H}_{T_r}(s)$ is then obtained as
\begin{align}
\hat{A}_r&=\hat{S}_r+\xi L_T,&& \hat{B}_{T,r}=\xi,&&\hat{C}_r&=C\hat{V}_{r,t}.\label{eq:18}
\end{align}
$\hat{H}_r(s)$ can be extracted from $\hat{H}_{T_r}(s)$ by setting $\hat{B}_r=-\hat{Q}_S^{-1}\hat{L}_r^*$. This particular choice of $\xi$ places the poles of $\hat{H}_{T_r}(s)$ (and $\hat{H}_r(s)$) at the mirror images of the interpolation points and makes the tangential directions their respective input-residuals. Thus, it enforces the interpolation condition (\ref{EQ44}) while also maintaining the structure of $\hat{B}_{T_r}$ according to (\ref{EQ42}). We now prove these properties in the following theorem, and we also show that the ROM $\hat{H}_r(s)$ obtained by enforcing these conditions is a time-limited pseudo-optimal ROM of $H(s)$.
\begin{theorem}If $(\hat{A}_r,\hat{B}_{T,r},\hat{C}_r)$ is defined as in the equation (\ref{eq:18}), then $\hat{H}_r(s)$, which is extracted from $\hat{H}_{T_r}(s)$ by setting $\hat{B}_r=-\hat{Q}_S^{-1}\hat{L}_r^*$, has the following properties:\\
(i) $\hat{H}_r(s)$ has poles at the mirror images of the interpolation points.\\
(ii) $\hat{c}_i$ is the input-residuals of the pole $-\sigma_i^*$ of $\hat{H}_r(s)$.\\
(iii) $\xi=\begin{bmatrix}\hat{B}_r&-e^{\hat{A}_rt}\hat{B}_r\end{bmatrix}$.\\
(iv) $\hat{Q}_S^{-1}$ is the time-limited controllability Gramian of the pair $(\hat{A}_r,\hat{B}_r)$.\\
(v) $\hat{H}_r(s)$ is a time-limited pseudo-optimal ROM of $H(s)$.
\end{theorem}
\textbf{Proof.} (i) By multiplying $\hat{Q}_S^{-1}$ from the left side of the equation (\ref{eq:17}) yields
\begin{align}
	-\hat{Q}_S^{-1}\hat{S}_r^*\hat{Q}_S-\hat{S}_r&+\hat{Q}_S^{-1}\hat{L}_r^*\hat{L}_r\nonumber\\
	-\hat{Q}_S^{-1}&e^{-\hat{S}_r^*t}\hat{L}_r^*\hat{L}_re^{-\hat{S}_rt}=0\nonumber\\
	-\hat{Q}_S^{-1}\hat{S}_r^*\hat{Q}_S&-\hat{A}_r=0.\nonumber
\end{align}
Thus, $\hat{A}_r=-\hat{Q}_S^{-1}\hat{S}_r^*\hat{Q}_S$, and hence, $\lambda_i(\hat{A}_r)=-\lambda_i(\hat{S}_r^*)$.\\
(ii) $(-\hat{Q}_S^{-1})(-\hat{S}_r^*)(-\hat{Q}_S)$ is actually the spectral factorization of $\hat{A}_r$. Also, $\hat{B}_r=-\hat{Q}_S^{-1}\begin{bmatrix}\hat{c}_i&\cdots&\hat{c}_r\end{bmatrix}^*$. Thus, $\hat{c}_i$ is the input-residual of $-\sigma_i^*$.\\
(iii) By putting the value of $\hat{B}_r$ in $\begin{bmatrix}\hat{B}_r&-e^{\hat{A}_rt}\hat{B}_r\end{bmatrix}$, we get $\begin{bmatrix}-\hat{Q}_S^{-1}\hat{L}_r^*&e^{\hat{A}_rt}\hat{Q}_S^{-1}\hat{L}_r^*\end{bmatrix}$. Since, $\hat{Q}_S e^{\hat{A}_rt}\hat{Q}_S^{-1}=e^{-\hat{S}_r^*t}$ and $e^{\hat{A}_rt}\hat{Q}_S^{-1}=\hat{Q}_S^{-1} e^{-\hat{S}_r^*t}$, $\xi=\begin{bmatrix}\hat{B}_r&-e^{\hat{A}_rt}\hat{B}_r\end{bmatrix}$.\\
(iv) The time-limited controllability Gramian of $(\hat{A}_r,\hat{B}_r)$ solves the following Lyapunov equation (as in the equation (\ref{eq:8b}))
\begin{align} \hat{A}_r\hat{P}_T+\hat{P}_T\hat{A}_r^*+\hat{B}_r\hat{B}_r^*-e^{\hat{A}_rt}\hat{B}_r\hat{B}^*e^{\hat{A}_r^*t}=0.\nonumber
\end{align}
By pre- and post-multiplying the equation (\ref{eq:8b}) with $\hat{Q}_S$, by putting $\hat{A}_r=-\hat{Q}_S^{-1}\hat{S}_r^*\hat{Q}_S$ and $\hat{B}_r=-\hat{Q}_S^{-1}\hat{L}_r^*$, and also by noting that $\hat{Q}_S e^{\hat{A}_rt}\hat{Q}_S^{-1}=e^{-\hat{S}_r^*t}$, the equation (\ref{eq:8b}) becomes
\begin{align}
	-\hat{S}_r^*\hat{Q}_S\hat{P}_T\hat{Q}_S&-\hat{Q}_S\hat{P}_T\hat{Q}_S\hat{S}_r\nonumber\\
	&+\hat{L}_r^*\hat{L}_r-e^{-\hat{S}_r^*t}\hat{L}_r^*\hat{L}_re^{-\hat{S}_rt}=0.\nonumber
\end{align}
Due to uniqueness, $\hat{Q}_S\hat{P}_T\hat{Q}_S=\hat{Q}_S$, $\hat{Q}_S\hat{P}_T=I$, and $\hat{P}_T=\hat{Q}_S^{-1}$.\\
(v) Consider the following equation
\begin{align}
	&A\hat{V}_{r,t}\hat{P}_T+\hat{V}_{r,t}\hat{P}_T\hat{A}_r^*+B\hat{B}_r^*-e^{At}B\hat{B}_r^*e^{\hat{A}_r^*t}\nonumber\\
	&=[\hat{V}_{r,t}\hat{S}_r+B_TL_T]\hat{P}_T-\hat{V}_{r,t}\hat{S}_r\hat{Q}_S^{-1}-B\hat{L}_r\hat{Q}_S^{-1}\nonumber\\
	&-e^{At}B\hat{L}_r\hat{Q}_S^{-1}e^{\hat{A}_r^*t}\nonumber\\
	&=\hat{V}_{r,t}\hat{S}_r\hat{P}_T+B\hat{L}_r\hat{P}_T-e^{At}B\hat{L}_re^{-\hat{S}_rt}\hat{P}_T-\hat{V}_{r,t}\hat{S}_r\hat{P}_T\nonumber\\
	&-B\hat{L}_r\hat{P}_T+e^{At}B\hat{L}_re^{-\hat{S}_rt}\hat{P}_T\nonumber\\
	&=0.\nonumber
\end{align}
Due to uniqueness, $\hat{V}_{r,t}\hat{P}_T=\tilde{P}_T$, and therefore, $\hat{C}_r\hat{P}_T=C\tilde{P}_T$. Hence, $\hat{H}_r(s)$ is a time-limited pseudo-optimal model ROM of $H(s)$.\\
\\We now aim to construct a ROM $\hat{G}_{T,r}(s)$, which satisfies the interpolation condition (\ref{EQ45}) while preserving the structure of $\hat{C}_{T,r}$ according to (\ref{EQ43}). Let $\sigma_k$ be the interpolation points in the (left) tangential directions $\hat{b}_k\in\mathbb{C}^{1\times p}$. Let us define $\bar{B}_T$ and $\tilde{B}_T$ as
\begin{align}
\bar{B}_T&=\begin{bmatrix}\hat{b}_1^*&\cdots&\hat{b}_r^*\end{bmatrix}^*, \nonumber\\ \tilde{B}_T&=\begin{bmatrix}\bar{B}_T&e^{-\hat{S}_rt}\bar{B}_T\end{bmatrix}=\begin{bmatrix}b_1^*&\cdots&b_r^*\end{bmatrix}^*.\label{eq:20}\end{align}
$\hat{G}_{T,r}(s)$ interpolates $G_T(s)$ at the interpolation points $\sigma_k$ in the respective (left) tangential directions $b_k\in\mathbb{C}^{2m\times1}$ for any input rational Krylov subspace $\hat{V}_{r,t}$ such that $\hat{W}_{r,t}^*\hat{V}_{r,t}=I$ if the output rational Krylov subspace $\hat{W}_{r,t}$ is defined as
\begin{align} \hat{W}_{r,t}=\begin{bmatrix}(A^*-\sigma_1I)^{-1}C_T^*b_1^*&\cdots&(A^*-\sigma_rI)^{-1}C_T^*b_r^*\end{bmatrix}\nonumber.
\end{align}
Owing to the relation with the Sylvester equation \cite{panzer2014model}, $\hat{W}_{r,t}$ solves the following Sylvester equation
\begin{align}
	\hat{W}_{r,t}^*A+(-\hat{S}_r)\hat{W}_{r,t}^*+(-\tilde{B}_T)(\tilde{L}_T)=0.\label{eq:21}
\end{align}
A family of ROMs which satisfy the interpolation condition $b_k^T\hat{G}_{T_r}(\sigma_k)=b_k^TG_T(\sigma_k)$ can be obtained by parametrizing $\hat{G}_{T_r}(\sigma_k)$ in $\hat{C}_{T,r}$ if all the interpolation points $\sigma_i$ have positive real parts, and the pair $(\hat{S}_r,\tilde{B}_T)$ is controllable, i.e.,
\begin{align}
	\hat{A}_r&=\hat{S}_r+\tilde{B}_T\xi, && \hat{B}_r=\hat{W}_{r,t}^*B,&&\hat{C}_{T,r}=\xi.\label{eq:22}
\end{align}
We now propose a novel choice of $\xi$ which retains the structure of $\hat{C}_{T_r}$ according to the equation (\ref{EQ43}), i.e.,
\begin{align}
	\xi=\begin{bmatrix}-\bar{B}_T^*\hat{P}_S^{-1}\\ \bar{B}_T^*e^{-\hat{S}_r^*}\hat{P}_S^{-1}\end{bmatrix},
\end{align} where $\hat{P}_S$ solves
\begin{align}
	-\hat{S}_r\hat{P}_S-\hat{P}_S \hat{S}_r^*+\bar{B}_T\bar{B}_T^*-e^{-\hat{S}_r}\bar{B}_T\bar{B}_T^*e^{-\hat{S}_r^*t}=0.\label{eq:27b}
\end{align}
$\hat{H}_r(s)$ can be extracted from $\hat{G}_{T_r}(s)$ by setting $\hat{C}_r=-\bar{B}_T^*\hat{P}_S^{-1}$. This particular choice of $\xi$ places the poles of $\hat{H}_r(s)$ at the mirror images of the interpolation points and makes the tangential directions their respective output-residuals. Thus, it enforces the interpolation condition (\ref{EQ45}) while also maintaining the structure of $\hat{C}_{T_r}$ according to (\ref{EQ43}). We now prove these properties in the following theorem, and we also show that the ROM $\hat{H}_r(s)$ obtained by enforcing these conditions is a time-limited pseudo-optimal ROM of $H(s)$.
\begin{theorem}If $(\hat{A}_r,\hat{B}_r,\hat{C}_{T,r})$ is defined as in the equation (\ref{eq:22}), then $\hat{H}_r(s)$, which is extracted from $\hat{G}_{T_r}(s)$ by setting $\hat{C}_r=-\bar{B}_T^*\hat{P}_S^{-1}$, has the following properties:\\
(i) $\hat{H}_r(s)$ has poles at the mirror images of the interpolation points.\\
(ii) $\hat{b}_i$ is the output residual of the pole $-\sigma_i^*$.\\
(iii) $\xi=\begin{bmatrix}\hat{C}_r \\ -\hat{C}_re^{\hat{A}_rt}\end{bmatrix}$.\\
(iv) $\hat{P}_S^{-1}$ is the time-limited observability Gramian of the pair $(\hat{A}_r,\hat{C}_r)$.\\
(v) $\hat{H}_r(s)$ is a time-limited pseudo-optimal ROM of $H(s)$.\end{theorem}
\textbf{Proof.} (i) By multiplying $\hat{P}_S^{-1}$ from the right, the equation (\ref{eq:27b}) becomes
\begin{align}
-\hat{S}_r-\hat{P}_S\hat{S}_r^*\hat{P}_S^{-1}+\bar{B}_T\bar{B}_T^*\hat{P}_S^{-1}&-e^{-\hat{S}_r}\bar{B}_T\bar{B}_T^*e^{-\hat{S}_r^*t}\hat{P}_S^{-1}=0\nonumber\\
-\hat{P}_S\hat{S}_r^*\hat{P}_S^{-1}-\hat{A}_r&=0.\nonumber
\end{align}
Thus, $\hat{A}_r=-\hat{P}_S\hat{S}_r^*\hat{P}_S^{-1}$, and hence, $\lambda_i(\hat{A}_r)=-\lambda_i(\hat{S}_r^*)$.\\
(ii) $(-\hat{P}_S)(-\hat{S}_r)(-\hat{P}_S^{-1})$ is actually the spectral factorization of $\hat{A}_r$. Moreover, $\hat{C}_r=\begin{bmatrix}\hat{b}_1^*&\cdots&\hat{b}_r^*\end{bmatrix}(-\hat{P}_S^{-1})$. Thus, $\hat{b}_i$ is the output residual of $-\sigma_i^*$.\\
(iii) By putting the value of $\hat{C}_r$ in $\begin{bmatrix}\hat{C}_r \\ -\hat{C}_re^{\hat{A}_rt}\end{bmatrix}$, we get $\begin{bmatrix}-\bar{B}_T^*\hat{P}_S^{-1} \\ \bar{B}_T^*\hat{P}_S^{-1}e^{\hat{A}_rt}\end{bmatrix}$. Since $\hat{P}_S^{-1} e^{\hat{A}_rt}\hat{P}_S=e^{-\hat{S}_r^*t}$ and $\hat{P}_S^{-1} e^{\hat{A}_rt}=e^{-\hat{S}_r^*t}\hat{P}_S^{-1}$, $\xi=\begin{bmatrix}\hat{C}_r \\ -\hat{C}_re^{\hat{A}_rt}\end{bmatrix}$.\\
(iv) The time-limited observability Gramian of $(\hat{A}_r,\hat{C}_r)$ solves the following Lyapunov equation (as in the equation (\ref{eq:9b}))
\begin{align} \hat{A}_r^*\hat{Q}_T+\hat{Q}_T\hat{A}_r+\hat{C}_r^*\hat{C}_r-e^{\hat{A}_r^*t}\hat{C}_r^*\hat{C}_re^{\hat{A}_rt}=0.\nonumber
\end{align}
By pre- and post-multiplying the equation (\ref{eq:9b}) with $\hat{P}_S$, by putting $\hat{A}_r=-\hat{P}_S\hat{S}_r^*\hat{P}_S^{-1}$ and $\hat{C}_r=-\bar{B}_T^*\hat{P}_S^{-1}$, and also by noting that $\hat{P}_S^{-1} e^{\hat{A}_rt}\hat{P}_S=e^{-\hat{S}_r^*t}$, the equation (\ref{eq:9b}) becomes
\begin{align}
	-\hat{S}_r\hat{P}_S\hat{Q}_T\hat{P}_S-\hat{P}_S\hat{Q}_T\hat{P}_S \hat{S}_r^*+\bar{B}_T\bar{B}_T^*\nonumber\\
-e^{-\hat{S}_r}\bar{B}_T\bar{B}_T^*e^{-\hat{S}_r^*t}=0.\nonumber
\end{align}
Due to uniqueness, $\hat{P}_S\hat{Q}_T\hat{P}_S=\hat{P}_S$, $\hat{P}_S\hat{Q}_T=I$, and $\hat{Q}_T=\hat{P}_S^{-1}$.\\
(v) Consider the following equation
\begin{align}
&\hat{A}_r^*\hat{Q}_T\hat{W}_{r,t}^*+\hat{Q}_T\hat{W}_{r,t}^*A+\hat{C}_r^*C-e^{\hat{A}_r^*t}\hat{C}_r^*Ce^{At}\nonumber\\
=&-\hat{Q}_T\hat{S}_r\hat{W}_{r,t}^*+\hat{Q_T}[\hat{S}_r\hat{W}_{r,t}^*+\tilde{B}_T\tilde{L}_T]\nonumber\\
&\hspace*{3.5cm}-\hat{Q}_T\bar{B}_TC+e^{\hat{A}_r^*t}\hat{Q}_T\bar{B}_TCe^{At}\nonumber\\
=&-\hat{Q}_T\hat{S}_r\hat{W}_{r,t}^*+\hat{Q_T}\hat{S}_r\hat{W}_{r,t}^*+\hat{Q}_T\bar{B}_TC-\hat{Q}_Te^{-\hat{S}_rt}\bar{B}_TCe^{At}\nonumber\\
&\hspace*{3.5cm}-\hat{Q}_T\bar{B}_TC+e^{\hat{A}_r^*t}\hat{Q}_T\bar{B}_TCe^{At}\nonumber\\
=&0\nonumber
\end{align}
Due to uniqueness, $\hat{Q}_T\hat{W}_{r,t}^*=\tilde{Q}_T$, and therefore, $\hat{Q}_T\hat{B}_r=\tilde{Q}_TB$. Hence, $\hat{H}_r(s)$ is a time-limited pseudo-optimal model ROM of $H(s)$.
\begin{remark}If the interpolation points are selected as the mirror images of $r$-poles of $H(s)$, TLPORK and O-TLPORK preserve these poles in $\hat{H}_r(s)$ like modal truncation. Additionally, TLPORK and O-TLPORK preserve the input and output residues, respectively of these poles if the tangential directions are selected as such.\end{remark}
\begin{remark}TLPORK and O-TLPORK resemble PORK in its pole-placement property, i.e., they place the poles at the mirror images of the interpolation points and make the tangential directions residuals of the poles. However, PORK does not (and does not have to) preserve a structure in the ROM. Unlike PORK, TLPORK and O-TLPORK do (and have to) preserve a specific structure in the ROM in order to satisfy the optimality conditions. When $t$ is set to $\infty$, TLPORK and O-TLPORK reduce to PORK.\end{remark}
\begin{remark}
TLPORK and O-TLPORK satisfy the optimality conditions (\ref{eq:9}) and (\ref{eq:10}), respectively, which are equivalent to (\ref{EQ32}) and (\ref{EQ33}), respectively. For SISO systems, the interpolation conditions (\ref{EQ32}) and (\ref{EQ33}) are equivalent. For MIMO systems, the interpolation conditions (\ref{EQ32}) and (\ref{EQ33}) are not equivalent. However, the ROMs constructed by TLPORK and O-TLPORK both satisfy (\ref{eq:13}). The accuracy of the ROMs in the MIMO case depends on the choice of tangential directions, even if the interpolation points used in TLPORK and O-TLPORK are the same.
\end{remark}
\begin{remark}The time-limited pseudo-optimality (equation (\ref{eq:13})) does not depend on the realization of $H(s)$ or $\hat{H}_r(s)$.
\end{remark}
\subsection{Algorithmic Aspects}
We allowed the state-space matrices to be complex so far in this section; however, one can obtain a real ROM for a real original model if the complex interpolation points (and the associated tangential directions) are chosen in conjugate pairs. This is a standard practice in Krylov subspace methods wherein the interpolation data is grouped into conjugate pairs to obtain a real basis (which satisfies the property in the equation (\ref{27c})); see \cite{grimme1997krylov} for more details. For instance, a real $\hat{V}_{r,t}$ can be computed by any rational Krylov subspace method instead of actually solving the Sylvester equation (\ref{eq:15}), i.e., \begin{align}
	Ran(\hat{V}_{r,t})=\underset {i=1,\cdots,r}{span}\{(\sigma_iI-A)^{-1}B_T c_i\}.\label{27c}
\end{align} The next step then is to compute the matrices of Sylvester equation which this $\hat{V}_{r,t}$ satisfies (as it may not satisfy the equation (\ref{eq:15})). This can be accomplished in a few simple steps. Choose any $\hat{W}_{r,t}$, for instance, $\hat{W}_{r,t}=\hat{V}_{r,t}$. Then compute the following matrices
\begin{align}
	\tilde{E}&=\hat{W}_{r,t}^T\hat{V}_{r,t}, \hspace*{0.3cm} \tilde{A}=\hat{W}_{r,t}^TA\hat{V}_{r,t},\hspace*{0.3cm}\tilde{B}=\hat{W}_{r,t}^TB_T,\label{28c}\\
	B_\bot&=B_T-\hat{V}_{r,t}\tilde{E}^{-1}\tilde{B}.
\end{align}
Then $L_T$ and $\hat{S}_r$ for the Sylvester equation of this $\hat{V}_{r,t}$ can be computed as
\begin{align}
	L_T&=(B_\bot^TB_\bot)^{-1}B_\bot^T\Big(A\hat{V}_{r,t}-\hat{V}_{r,t}\tilde{E}^{-1}\tilde{A}\Big),\\
	\hat{S}_r&=\tilde{E}^{-1}\Big(\tilde{A}-\tilde{B}L_T\Big).\label{31c}
\end{align}
Now partition $L_T$ as $L_T=\begin{bmatrix}L_{m\times r}^+\\L_{m\times r}^-\end{bmatrix}$, and define $L_T^-$ as $L_T^-=\begin{bmatrix}L_{m\times r}^+\\-L_{m\times r}^-\end{bmatrix}$. Then $\hat{Q}_S$ can be computed from the following Lyapunov equation
\begin{align}
	-\hat{S}_r^T\hat{Q}_S-\hat{Q}_S\hat{S}_r+L_T^TL_T^-=0.\label{32c}
\end{align}
Finally, the ROM is obtained as
\begin{align}
\hat{A}_r&=-\hat{Q}_S^{-1}\hat{S}_r^T\hat{Q}_S, &&\hat{B}_r=-\hat{Q}_S^{-1}(L_{m\times r}^+)^T,\nonumber\\
\hat{C}_r&=C\hat{V}_{r,t}.\label{33d}\end{align}
\textbf{Algorithm 1: TLPORK}\\
\textbf{Input:} Original model: $(A,B,C)$, interpolation points: $\{\sigma_{1},\cdots,\sigma_{r}\}$, tangential directions: $\{\hat{c}_1,\cdots,\hat{c}_{r}\}$.\\
\textbf{Output:} ROM: $(\hat{A}_r,\hat{B}_r,\hat{C}_r)$.\\
(i) Define $B_T$ according to the equation (\ref{EQ42}), and $\hat{S}_r$ and $L_T$ according to the equation (\ref{eq:14}).\\
(ii) Compute $\hat{V}_{r,t}$ from the equation (\ref{27c}).\\
(iii) Update $\hat{S}_r$ and $L_T$ from the equations (\ref{28c})-(\ref{31c}).\\
(iv) Partition $L_T$ as $L_T=\begin{bmatrix}L_{m\times r}^+&L_{m\times r}^-\end{bmatrix}^T$.\\
(v) Define $L_T^-=\begin{bmatrix}L_{m\times r}^+&-L_{m\times r}^-\end{bmatrix}^T$.\\
(vi) Solve the equation (\ref{32c}) to calculate $\hat{Q}_S$.\\
(vii) Obtain the ROM from the equation (\ref{33d}).\\
\\Similarly, a real $\hat{W}_{r,t}$ can be computed by any rational Krylov subspace method instead of actually solving the Sylvester equation (\ref{eq:21}) if the complex interpolation points (and the associated tangential directions) are chosen in conjugate pairs, i.e.,
\begin{align}
	Ran(\hat{W}_{r,t})&=\underset {i=1,\cdots,r}{span}\{(\sigma_iI-A^T)^{-1}C_T^T b_i^T\}.\label{33c}
\end{align}
The next step then is to compute the matrices of Sylvester equation which this $\hat{W}_{r,t}$ satisfies (as it may not satisfy the equation (\ref{eq:21})). Choose any $\hat{V}_{r,t}$, for instance, $\hat{V}_{r,t}=\hat{W}_{r,t}$. Then compute the following matrices
\begin{align}
	\tilde{E}&=\hat{W}_{r,t}^T\hat{V}_{r,t}, \hspace*{0.3cm}\tilde{A}=\hat{W}_{r,t}^TA\hat{V}_{r,t},\hspace*{0.3cm}\tilde{C}=C_T\hat{V}_{r,t},\label{34c}\\
	C_\bot&=\tilde{L}_T-\tilde{C}\tilde{E}^{-1}\hat{W}_{r,t}^T.
\end{align}
Then $\tilde{B}_T$ and $\hat{S}_r$ for the Sylvester equation of this $\hat{W}_{r,t}$ can be computed as
\begin{align}
	\tilde{B}_T&=\Big(\hat{W}_{r,t}^TA-\tilde{A}\tilde{E}^{-1}\hat{W}_{r,t}^T\Big)C_\bot^T(C_\bot C_\bot^T)^{-1},\\
	\hat{S}_r&=\Big(\tilde{A}-\tilde{B}_T\tilde{C}\Big)\tilde{E}^{-1}.\label{37c}
\end{align}
Now partition $\tilde{B}_T$ as $\tilde{B}_T=\begin{bmatrix}\tilde{B}_{m\times r}^+&\tilde{B}_{m\times r}^-\end{bmatrix}$, and define $\tilde{B}_T^-$ as $\tilde{B}_T^-=\begin{bmatrix}\tilde{B}_{m\times r}^+&-\tilde{B}_{m\times r}^-\end{bmatrix}^T$. Then $\hat{P}_S$ can be computed from the following Lyapunov equation
\begin{align}
	-\hat{S}_r\hat{P}_S-\hat{P}_S\hat{S}_r^T+\tilde{B}_T\tilde{B}_T^-=0.\label{38c}
\end{align}
The ROM is obtained as
\begin{align}\hat{A}_r&=-\hat{P}_S\hat{S}_r^T\hat{P}_S^{-1}, &&\hat{B}_r=\hat{W}_{r,t}^{T}B,\nonumber\\
\hat{C}_r&=-(\tilde{B}_{m\times r}^+)^T\hat{P}_S^{-1}.\label{40d}\end{align}
\textbf{Algorithm 2: O-TLPORK}\\
\textbf{Input:} Original model: $(A,B,C)$, interpolation points: $\{\sigma_{1},\cdots,\sigma_{r}\}$, tangential directions: $\{\hat{b}_1,\cdots,\hat{b}_{r}\}$.\\
\textbf{Output:}  ROM: $(\hat{A}_r,\hat{B}_r,\hat{C}_r)$.\\
(i) Define $C_T$ according to the equation (\ref{EQ43}), and $\hat{S}_r$ and $\tilde{B}_T$ according to the equation (\ref{eq:20}).\\
(ii) Compute $\hat{W}_{r,t}$ from the equation (\ref{33c}).\\
(iii) Update $\hat{S}_r$ and $\tilde{B}_T$ from the equations (\ref{34c})-(\ref{37c}).\\
(iv) Partition $\tilde{B}_T$ as $\tilde{B}_T=\begin{bmatrix}\tilde{B}_{m\times r}^+&\tilde{B}_{m\times r}^-\end{bmatrix}$.\\
(v) Define $\tilde{B}_T^-$ as $\tilde{B}_T^-=\begin{bmatrix}\tilde{B}_{m\times r}^+&-\tilde{B}_{m\times r}^-\end{bmatrix}^T$.\\
(vi) Solve the equation (\ref{38c}) to calculate $\hat{P}_S$.\\
(vii) Obtain the ROM from the equation (\ref{40d}).
\subsection{TLCURE}
We now generalize CURE for the time-limited scenario aiming to inherit the monotonic decay in the error property when TLPORK/O-TLPORK is used within CURE. Applying CURE on $H_T(s)$ and $G_T(s)$ can generate their ROMs $H_{T_r}(s)$ and $G_{T_r}(s)$, respectively with a monotonic decay in the error. However, the ROM $\hat{H}_r(s)$ cannot be extracted from $H_{T_r}(s)$ or $G_{T_r}(s)$ because the CURE framework does not preserve the structure according to the equation (\ref{EQ42}) or (\ref{EQ43}). This is the same issue that TLIRKA faces. In the sequel, we tackle this issue and generalize CURE to adaptively construct the ROMs whose $\mathcal{H}_{2,t}$-norm error decay monotonically irrespective of the choice of interpolation points and tangential directions.\\
Recall the ROM constructed by PORK is given by
\begin{align}
\hat{A}_r=-\hat{Q}_s^{-1}\hat{S}_r^T\hat{Q}_s,&& B_r=-\hat{Q}_s\hat{L}_r^T,&&\hat{C}_r=C\hat{V}_r.\nonumber
\end{align}
Since pseudo-optimality is a property of the transfer function, and it does not depend on the state-space realization, a state transformation can be applied using $\hat{Q}_s^{-1}$ as the transformation matrix, i.e.,
\begin{align}
\hat{A}_r=-\hat{S}_r^T,&& \hat{B}_r=-\hat{L}_r^T,&&\hat{C}_r=C\hat{V}_r\hat{Q}_s^{-1}.\nonumber
\end{align}
On similar lines, the ROM constructed by TLPORK can be defined as
\begin{align}
\hat{A}_r=-\hat{S}_r^T,&& \hat{B}_r=-\hat{L}_r^T,&&\hat{C}_r=C\hat{V}_{r,t}\hat{Q}_S^{-1}.\nonumber
\end{align}
Since $\hat{Q}_S$ satisfies the equation (\ref{32c}), the following holds (see \cite{gawronski1990model} for details)
\begin{align}
\hat{Q}_S&=\hat{Q}_s-e^{-\hat{S}_r^Tt}\hat{Q}_se^{-\hat{S}_rt}.
\end{align}
Similarly, the following relationship holds between $\hat{V}_{r,t}$ and $\hat{V}_r$, i.e.,
\begin{align}
\hat{V}_{r,t}&=\hat{V}_r-e^{At}\hat{V}_re^{-\hat{S}_rt}.\label{Eq:47A}
\end{align} This can be verified as the following by noting that $Ae^{At}=e^{At}A$ and $\hat{S}_re^{\hat{S}_rt}=e^{\hat{S}_rt}\hat{S}_r$, i.e.,
\begin{align}
A\hat{V}_{r,t}&=A\hat{V}_r-Ae^{At}\hat{V}_re^{-\hat{S}_rt}\nonumber\\
A\hat{V}_{r,t}&=A\hat{V}_r-e^{At}A\hat{V}_re^{-\hat{S}_rt}\nonumber\\
A\hat{V}_{r,t}&=\hat{V}_r\hat{S}_r+B\hat{L}_r-e^{At}\big(\hat{V}_r\hat{S}_r+B\hat{L}_r\big)e^{-\hat{S}_rt}\nonumber\\
A\hat{V}_{r,t}&=\hat{V}_r\hat{S}_r+B\hat{L}_r-e^{At}\hat{V}_r\hat{S}_re^{-\hat{S}_rt}-e^{At}B\hat{L}_re^{-\hat{S}_rt}\nonumber\\
A\hat{V}_{r,t}&=\big(\hat{V}_r-e^{At}\hat{V}_re^{-\hat{S}_rt}\big)\hat{S}_r+B\hat{L}_r-e^{At}B\hat{L}_re^{-\hat{S}_rt}\nonumber\\
A\hat{V}_{r,t}&=\hat{V}_{r,t}\hat{S}_r+B\hat{L}_r-e^{At}B\hat{L}_re^{-\hat{S}_rt}.\nonumber
\end{align}
Thus the time-limited pseudo-optimality can be enforced on the pair $(\hat{A}_r,\hat{B}_r)$ generated by PORK by selecting $\hat{C}_r$ as $C\hat{V}_{r,t}\hat{Q}_S$.\\
It is shown in \cite{wolf2014h} that $P_{tot}^{(i)}$ can be obtained recursively if $(\hat{A}^{(i)},\hat{B}^{(i)},\hat{C}^{(i)})$ is obtained at each step of CURE using PORK, i.e.,
\begin{align}
P_{tot}^{(i)}=\begin{bmatrix}P_{tot}^{(i-1)}&0\\0&\big(\hat{Q}^{(i)}\big)^{-1}\end{bmatrix}\label{EQ72}
\end{align} where
\begin{align}
-(\hat{S}^{(i)})^T\hat{Q}^{(i)}-\hat{Q}^{(i)}\hat{S}^{(i)}+(\hat{L}^{(i)})^T\hat{L}^{(i)}=0.\label{EQ73}
\end{align}
One can note that $S_{tot}^{(i)}$ and $L_{tot}^{(i)}$ do not depend on $C_{tot}^{(i)}$ or $C_{tot}^{(i-1)}$ in each step of CURE. Thus the basic structure of CURE is not affected if $C_{tot}^{(i)}$ is manipulated in each step. We now show that if the time-limited pseudo-optimality is enforced at each step of CURE by appropriately constructing $C_{tot}^{(i)}$, $||H(s)-H_{tot}^{(i)}\big(s\big)||^2_{\mathcal{H}_{2,t}}$ decays monotonically at each step irrespective of the choice of interpolation points and the tangential directions.\\
Let $(\hat{A}^{(i)},\hat{B}^{(i)},\hat{C}^{(i)})$ is constructed using PORK in every iteration of CURE, and let $H_{tot}^{(i)}\big(s\big)$ is constructed as the following
\begin{align}
A_{tot}^{(i)}&=(-S_{tot}^{(i)})^T,&& B_{tot}^{(i)}=(-L_{tot}^{(i)})^T,\nonumber\\ C_{tot}^{(i)}&=CV_{tot,t}^{(i)}P_{tot,t}^{(i)}\label{EQ74}
\end{align} where
\begin{align}
V_{tot,t}^{(i)}&=V_{tot}^{(i)}-e^{At}V_{tot}^{(i)}e^{-S_{tot}^{(i)}t},\label{EQ75}\\
P_{tot,t}^{(i)}&=\Big(\big(P_{tot}^{(i)}\big)^{-1}-e^{(-S_{tot}^{(i)})^Tt}\big(P_{tot}^{(i)}\big)^{-1}e^{-S_{tot}^{(i)}t}\Big)^{-1}.\label{EQ76}
\end{align}
Needless to say that $H_{tot}^{(i)}\big(s\big)$ is time-limited pseudo-optimal ROM of $H(s)$ as the time-limited pseudo-optimality is judiciously enforced on $H_{tot}^{(i)}\big(s\big)$ by setting $C_{tot}^{(i)}=CV_{tot,t}^{(i)}P_{tot,t}^{(i)}$. Thus
\begin{align}
||H(s)-H_{tot}^{(i)}\big(s\big)||^2_{\mathcal{H}_{2,t}}&=||H(s)||^2_{\mathcal{H}_{2,t}}-||H_{tot}^{(i)}\big(s\big)||^2_{\mathcal{H}_{2,t}}.\nonumber
\end{align}
Moreover, $H_{tot}^{(i-1)}\big(s\big)$ is a time-limited pseudo-optimal ROM of $H_{tot}^{(i)}\big(s\big)$ as $\big(S_{tot}^{(i)},L_{tot}^{(i)}\big)$ contains the interpolation points and tangential directions of the pair $\big(S_{tot}^{(i-1)},L_{tot}^{(i-1)}\big)$. This is because the basic structure of CURE is not affected due to the choice of $H_{tot}^{(i)}\big(s\big)$ according to the equation (\ref{EQ74}). Therefore,
\begin{align}
||H_{tot}^{(i)}\big(s\big)&-H_{tot}^{(i-1)}\big(s\big)||^2_{\mathcal{H}_{2,t}}\nonumber\\
&=||H_{tot}^{(i)}\big(s\big)||^2_{\mathcal{H}_{2,t}}-||H_{tot}^{(i-1)}\big(s\big)||^2_{\mathcal{H}_{2,t}},\nonumber\\
||H_{tot}^{(i)}\big(s\big)||^2_{\mathcal{H}_{2,t}}&\geq ||H_{tot}^{(i-1)}\big(s\big)||^2_{\mathcal{H}_{2,t}}.\nonumber
\end{align}
Hence, $||H(s)-H_{tot}^{(i)}\big(s\big)||^2_{\mathcal{H}_{2,t}}$ decays monotonically in each step irrespective of the choice of interpolation points and tangential directions. Also, note that
\begin{align}
||H(s)-H_{tot}^{(i)}\big(s\big)||^2_{\mathcal{H}_{2,t}}&=CP_TC^T-CV_{tot,t}^{(i)}P_{tot,t}^{(i)}(V_{tot,t}^{(i)})^TC^T\nonumber\\
&=C\Big(P_T-V_{tot,t}^{(i)}P_{tot,t}^{(i)}(V_{tot,t}^{(i)})^T\Big)C^T.\nonumber
\end{align} Therefore, $V_{tot,t}^{(i)}P_{tot,t}^{(i)}(V_{tot,t}^{(i)})^T$ monotonically approaches to $P_T$ after each iteration of TLCURE. Hence, TLCURE also provides an approximation of $P_T$.\\
\\\textbf{Algorithm 3: TLCURE (V-type)}\\
(i) Initialize $B_{\perp}^{(0)}=B$, $S_{tot}^{(0)}=[\hspace{2mm}]$, $C_{tot}^{(0)}=[\hspace{2mm}]$, $L_{tot}^{(0)}=[\hspace{2mm}]$, $P_{tot}^{(0)}=[\hspace{2mm}]$, $\bar{B}_{tot}^{(0)}=[\hspace{2mm}]$, $V_{tot}^{(i)}=[\hspace{2mm}]$.\\
\textbf{for} $i=1,\cdots,k$\\
(ii) $Ran(\hat{V}^{(i)})=\underset {j=1,\cdots,r_j}{span}\{(\sigma_jI-A)^{-1}Bb_j\}$.\\
(iii) Set $\hat{W}^{(i)}=\hat{V}^{(i)}(\hat{V}^{(i)}(\hat{V}^{(i)})^T)^{-1}$, $\hat{A}=(\hat{W}^{(i)})^TA\hat{V}^{(i)}$,\\
\hspace*{4.2mm}$\hat{B}=(\hat{W}^{(i)})^TB_\bot^{(i-1)}$, $B_\bot^{(i)}=B_\bot^{i-1}-\hat{V}^{(i)}\hat{B}$,\\ \hspace*{4.2mm}$\hat{L}^{(i)}=((B_\bot^{(i)})^TB_\bot^{(i)})^{-1}(B_\bot^{(i)})^T(A\hat{V}^{(i)}-\hat{V}^{(i)}\hat{A})$,\\ \hspace*{4.2mm}$\hat{S}^{(i)}=\hat{A}-\hat{B}\hat{L}^{(i)}$.\\
(iv) Solve the equation (\ref{EQ73}) to compute $\hat{Q}^{(i)}$.\\
(v) Set $P_{tot}^{(i)}$ according to the equation (\ref{EQ72}).\\
(vi) $\bar{B}_{tot}^{(i)}=\begin{bmatrix}\bar{B}_{tot}^{(i-1)}\\-(\hat{Q}^{(i)})^{-1}(\hat{L}^{(i)})^T\end{bmatrix}$, $L_{tot}^{(i)}=\begin{bmatrix}L_{tot}^{(i-1)}&\hat{L}^{(i)}\end{bmatrix}$, \\ $S_{tot}^{(i)}=\begin{bmatrix}S_{tot}^{(i-1)}&-\bar{B}_{tot}^{(i-1)}\hat{L}^{(i)}\\0&\hat{S}^{(i)}\end{bmatrix}$, $V_{tot}^{(i)}=\begin{bmatrix}V_{tot}^{(i-1)}&\hat{V}^{(i)}\end{bmatrix}$.\\
(vii) Compute $V_{tot,t}^{(i)}$ and $P_{tot,t}^{(i)}$ from the equations (\ref{EQ75}) and (\ref{EQ76}), respectively.\\
(viii) Compute cumulative ROM from the equation (\ref{EQ74}).\\
\textbf{end for}.\\
\\We have so far just presented a V-type algorithm, i.e., only $V_{tot,t}^{(i)}$ is computed for the construction of the ROM. A W-type algorithm can also be developed by duality, which we now present without a proof for brevity. The W-type TLCURE is summarized in algorithm $4$. It also ensures that $||H(s)-H_{tot}^{(i)}\big(s\big)||^2_{\mathcal{H}_{2,t}}$ decays monotonically at each step irrespective of the choice of interpolation points and the tangential directions provided $H_{tot}^{(i)}$ is computed using O-TLPORK at each step.\\
\\\textbf{Algorithm 4: TLCURE (W-type)}\\
(i) Initialize $C_{\perp}^{(0)}=C$, $S_{tot}^{(0)}=[\hspace{2mm}]$, $C_{tot}^{(0)}=[\hspace{2mm}]$, $L_{tot}^{(0)}=[\hspace{2mm}]$, $Q_{tot}^{(0)}=[\hspace{2mm}]$, $\bar{C}_{tot}^{(0)}=[\hspace{2mm}]$, $W_{tot}^{(i)}=[\hspace{2mm}]$.\\
\textbf{for} $i=1,\cdots,k$\\
(ii) $Ran(\hat{W}^{(i)})=\underset {j=1,\cdots,r_j}{span}\{(\sigma_jI-A^T)^{-1}C^Tc_j^T\}$.\\
(iii) Set $\hat{V}^{(i)}=\hat{W}^{(i)}$, $\hat{W}^{(i)}=\hat{W}^{(i)}(\hat{W}^{(i)}(\hat{W}^{(i)})^T)^{-1}$,\\ \hspace*{4.2mm}$\hat{A}=(\hat{W}^{(i)})^TA\hat{V}^{(i)}$, $\hat{C}=C_\bot^{(i-1)}\hat{V}^{(i)}$,\\ \hspace*{4.2mm}$C_\bot^{(i)}=C_\bot^{i-1}-\hat{C}(\hat{W}^{(i)})^T$,\\ \hspace*{4.2mm}$\hat{L}^{(i)}=\big((\hat{W}^{(i)})^TA-\hat{A}(\hat{W}^{(i)})^T\big)(C_\bot^{(i)})^T\big(C_\bot^{(i)}(C_\bot^{(i)})^T\big)^{-1}$,\\
\hspace*{4.2mm}$\hat{S}^{(i)}=\hat{A}-\hat{L}^{(i)}\hat{C}$.\\
(iv) Solve $-S^{(i)}\hat{P}^{(i)}-\hat{P}^{(i)}(\hat{S}^{(i)})^T+\hat{L}^{(i)}(\hat{L}^{(i)})^T=0$.\\
(v) $Q_{tot}^{(i)}=\begin{bmatrix}Q_{tot}^{(i-1)}&0\\0&(\hat{P}^{(i)})^{-1}\end{bmatrix}$, $\bar{C}_{tot}^{(i)}=\begin{bmatrix}\bar{C}_{tot}^{(i-1)}\\\\-\big(\hat{L}^{(i)}\big)^T(\hat{P}^{(i)})^{-1}\end{bmatrix}$, \hspace*{4.2mm}$S_{tot}^{(i)}=\begin{bmatrix}S_{tot}^{(i-1)}&0\\-\hat{L}^{(i)}\bar{C}_{tot}^{(i-1)}&\hat{S}^{(i)}\end{bmatrix}$, $L_{tot}^{(i)}=\begin{bmatrix}L_{tot}^{(i-1)}\\\hat{L}^{(i)}\end{bmatrix}$,\\ \hspace*{4.2mm}$W_{tot}^{(i)}=\begin{bmatrix}W_{tot}^{(i-1)}& \hat{W}^{(i)}\end{bmatrix}$.\\
(vi) $Q_{tot,t}^{(i)}=\Big((Q_{tot}^{(i)})^{-1}-e^{-S_{tot}^{(i)}t}(Q_{tot}^{(i)})^{-1}e^{(-S_{tot}^{(i)})^Tt}\Big)^{-1}$.\\
(vii) $W_{tot,t}^{(i)}=W_{tot}^{(i)}-e^{A^Tt}W_{tot}^{(i)}e^{(-S_{tot}^{(i)})^Tt}$.\\
(viii) $A_{tot}^{(i)}=(-S_{tot}^{(i)})^T$, $C_{tot}^{(i)}=(-L_{tot}^{(i)})^T$,\\ \hspace*{6mm}$B_{tot}^{(i)}=Q_{tot,t}^{(i)}(W_{tot,t}^{(i)})^TB$.\\
\textbf{end for}.
\begin{remark}
It should be stressed here that it does not matter whether a time-limited pseudo-optimal ROM is generated in a single run of TLPORK (,i.e., algorithm $1$ or $2$) or adaptively using TLCURE (i.e., algorithm $3$ or $4$) as long as the same interpolation points and the tangential directions are used. The real benefit of TLCURE and its monotonic decay in the error property is that it makes an adaptive choice of the order $r$ of the ROM possible for a given tolerance in terms of $\mathcal{H}_{2,t}$-norm error.
\end{remark}
Note that
\begin{align}
||H(s)&-H_{tot}^{(i)}\big(s\big)||^2_{\mathcal{H}_{2,t}}\nonumber\\
&=B^TQ_TB-B^TW_{tot,t}^{(i)}Q_{tot,t}^{(i)}(W_{tot,t}^{(i)})^TB\nonumber\\
&=B^T\big(Q_T-W_{tot,t}^{(i)}Q_{tot,t}^{(i)}(W_{tot,t}^{(i)})^T\big)B.\nonumber
\end{align} Therefore, $W_{tot,t}^{(i)}Q_{tot,t}^{(i)}(W_{tot,t}^{(i)})^T$ monotonically approaches to $Q_T$ after each iteration of TLCURE (W-type).
\subsection{Extension to Generalized Time-limited Problem}
The results presented so far can readily be generalized for any time interval other than $[0,t]$, i.e., $[t_1,t_2]$ where $0<t_1<t_2$. This can be done by making some simple changes in the algorithms $1-4$, which we now enlist. In the algorithm $1$, $B_T$ and $L_T$ in the step (1) are replaced with the following:
\begin{align}
B_T=\begin{bmatrix}e^{At_1}B & -e^{At_2}B\end{bmatrix}\textnormal{ and }L_T=\begin{bmatrix}\hat{L}_re^{-\hat{S}t_1}\\\hat{L}_re^{-\hat{S}t_2}\end{bmatrix}.\nonumber
\end{align}
In the algorithm $2$, $C_T$ and $\tilde{B}_T$ in the step (1) are replaced with the following:
\begin{align}
C_T=\begin{bmatrix}Ce^{At_1} \\ -Ce^{At_2}\end{bmatrix}\textnormal{ and }\tilde{B}_T=\begin{bmatrix}e^{-\hat{S}t_1}\bar{B}_T&e^{-\hat{S}t_2}\bar{B}_T\end{bmatrix}.\nonumber
\end{align}
In the algorithm $3$, $V_{tot,t}^{(i)}$ and $P_{tot,t}^{(i)}$ in the step (8) are replaced with the following:
\begin{align}
V_{tot,t}^{(i)}&=e^{At_1}V_{tot}^{(i)}e^{-S_{tot}^{(i)}t_1}-e^{At_2}V_{tot}^{(i)}e^{-S_{tot}^{(i)}t_2} \hspace*{1cm}\textnormal{and}\nonumber\\
P_{tot,t}^{(i)}&=\Big(e^{(-S_{tot}^{(i)})^Tt_1}\big(P_{tot}^{(i)}\big)^{-1}e^{-S_{tot}^{(i)}t_1}\nonumber\\
&\hspace*{2.5cm}-e^{(-S_{tot}^{(i)})^Tt_2}\big(P_{tot}^{(i)}\big)^{-1}e^{-S_{tot}^{(i)}t_2}\Big)^{-1}.\nonumber
\end{align}
In the algorithm $4$, $Q_{tot,t}^{(i)}$ and $W_{tot,t}^{(i)}$ in the steps (7) and (8), respectively are replaced with the following:
\begin{align}
Q_{tot,t}^{(i)}&=\Big(e^{-S_{tot}^{(i)}t_1}\big(Q_{tot}^{(i)}\big)^{-1}e^{(-S_{tot}^{(i)})^Tt_1}\nonumber\\
&\hspace*{2cm}-e^{-S_{tot}^{(i)}t_2}\big(Q_{tot}^{(i)}\big)^{-1}e^{(-S_{tot}^{(i)})^Tt_2}\Big)^{-1}\textnormal{ and}\nonumber\\
W_{tot,t}^{(i)}&=e^{A^Tt_1}W_{tot}^{(i)}e^{(-S_{tot}^{(i)})^Tt_1}-e^{A^Tt_2}W_{tot}^{(i)}e^{(-S_{tot}^{(i)})^Tt_2}.\nonumber
\end{align}
\subsection{Choice of the User-defined Parameters}
There is no theoretical assurance that the particular choices of the interpolation points and the tangential directions ensure the least error. However, some general guidelines can be followed to construct a high-fidelity ROM. For instance, it is suggested in \cite{gugercin2008h_2} to interpolate at the mirror images of the poles of the original system, which are associated with the large residuals to obtain less error. The preservation of poles associated with the large residuals in itself generally leads to a good time- and frequency-domain accuracy \cite{rommes2006efficient}. Unlike TLIRKA, the poles of the ROM $\hat{H}_r(s)$ are known a priori in TLPORK and O-TLPORK, and it can preserve the desired poles of the original system as well. Moreover, TLPORK and O-TLPORK can also preserve the input and output residuals, respectively, associated with these poles. These properties can be exploited to obtain a high-fidelity ROM while satisfying a subset of the optimality conditions at the same time. Moreover, TLPORK and O-TLPORK can be adaptively constructed using the TLCURE framework, and the order of the ROM can be increased until the error has decayed up to the desired accuracy. Thus, the adaptive nature of TLCURE can be exploited to automatically select the order of the ROM for the desired tolerance in the accuracy. The computation of $\mathcal{H}_{2,t}$-norm can become a computationally demanding task in a large-scale setting as it requires the computation of $P_T$ or $Q_T$. In this scenario, $P_T$ or $Q_T$ may be replaced with its low-rank approximation, as suggested in \cite{kurschner2018balanced,redmann2017mathcal}.
\subsection{Computational Cost}
TLBT \cite{gawronski1990model} is computationally not feasible for large-scale systems because of the cubic complexity associated with the solution of large-scale Lyapunov equations. In \cite{kurschner2018balanced}, the applicability of TLBT is extended to large-scale systems by using the low-rank approximation of Lyapunov equations. Various methods to efficiently compute $e^{At}B$ and $Ce^{At}$ are also discussed in \cite{kurschner2018balanced}. TLPORK and O-TLPORK do not involve large-scale Lyapunov equations like TLBT \cite{gawronski1990model}. The Lyapunov equations involved in the algorithm, i.e., the equations (\ref{32c}) and (\ref{38c}) are small-scale equations, which can be solved easily without much computational effort. The main computational effort in TLPORK and O-TLPORK is spent on calculating the rational Krylov subspaces $\hat{V}_{r,t}$ and $\hat{W}_{r,t}$ for which several efficient methods are available \cite{panzer2014model}. Thus TLPORK and O-TLPORK are easily applicable to large-scale systems.
\section{Numerical Results}
In this section, we perform three experiments to test the efficacy of TLPORK, and we compare its performance with the well-known existing techniques. The first two test models are taken from the benchmark collection of \cite{chahlaoui2002collection}, and the third model is taken from \cite{rommes2009computing}. In all the experiments, we initialize IRKA randomly and use its final interpolation points and tangential directions to initialize ISRKA and TLIRKA. We use the same interpolation points and tangential directions for PORK. We use the final interpolation points and tangential directions of TLIRKA for TLPORK and O-TLPORK. This helps to investigate the impact of satisfying only a subset of the optimality conditions on the accuracy of the ROM. The results of O-TLPORK are indistinguishable from TLPORK for the SISO case and hence, not shown in the figures for clarity. The approximate Gramians provided by TLPORK and O-TLPORK, i.e., $\hat{V}_{r,t}\hat{P}_T\hat{V}_{r,t}^T$ and $\hat{W}_{r,t}\hat{Q}_T\hat{W}_{r,t}^T$, respectively are used to implement TLBT, and we refer it to here as ``Approximate-TLBT (A-TLBT)". The $\mathcal{H}_{2,t}$-norm of the error is computed from the equations (\ref{EQ7})-(\ref{EQ12}). We solved the large-scale Lyapunov equations (\ref{EQ7})-(\ref{EQ12}) exactly for the sake of accuracy and the fairness of the comparison. Practically, however, these can be replaced with their low-rank approximations, as suggested in \cite{redmann2017mathcal}. All the experiments are conducted on a computer with Intel(R) Core(TM) i7-8550U 1.80GHz$\times$8 processors and 16GB memory using MATLAB $2016$.

\begin{figure}[!h]
\centering
\includegraphics[width=8.2cm]{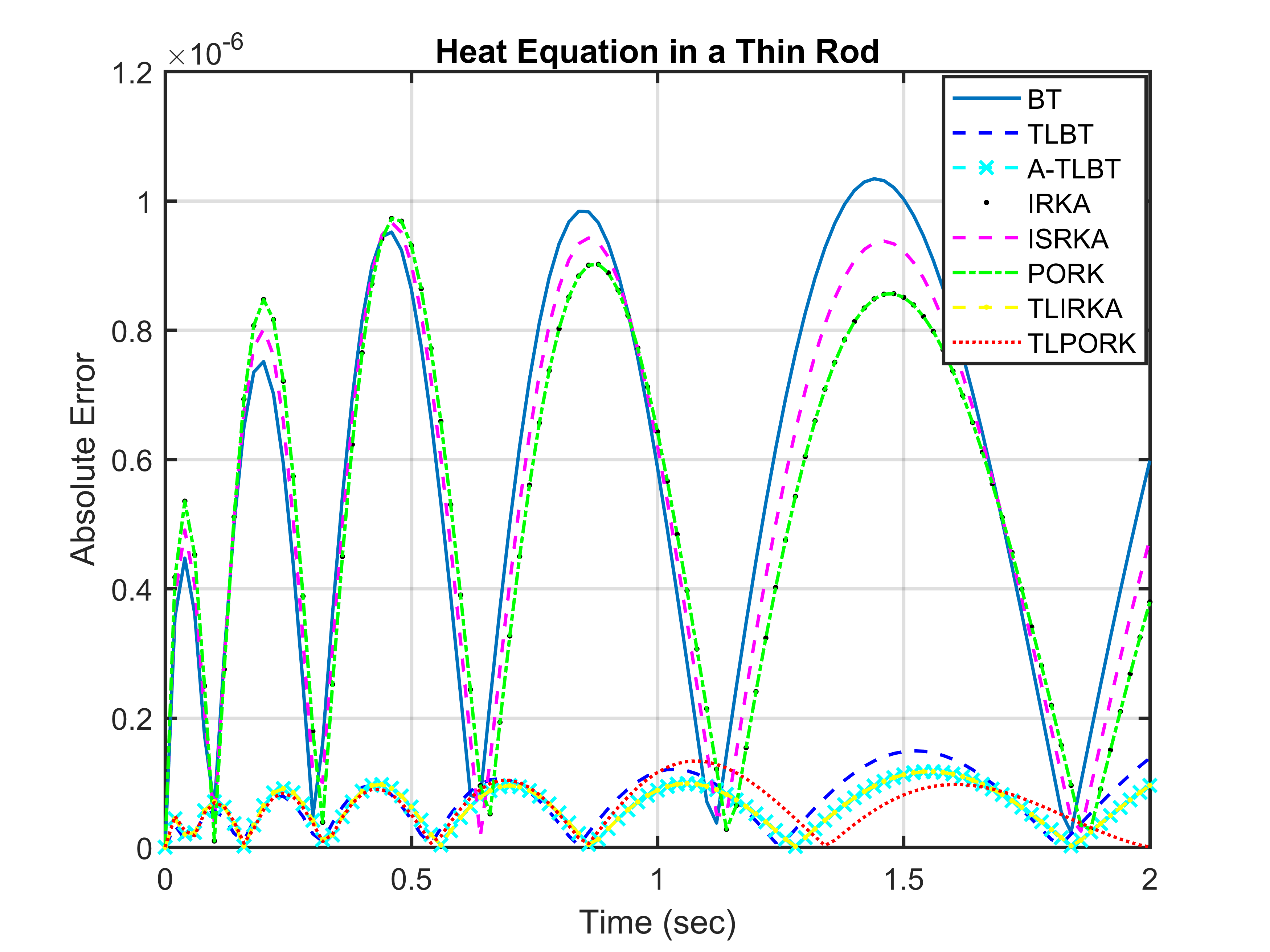}
\caption{Absolute error in the step response}\label{fig:1}
\end{figure}

\textbf{Heat equation in a thin rod:} This is a $200^{th}$ order SISO model taken from \cite{chahlaoui2002collection}. In this experiment, a unit step input is applied to the system at $0$ sec for two seconds. The goal of the experiment is to construct a ROM that accurately mimics the original system during these two seconds when the same input is applied. A $5^{th}$ order ROM is obtained using BT, TLBT, IRKA, ISRKA, PORK, TLIRKA, TLPORK, O-TLPORK, and A-TLBT. The desired time interval is specified as $[0,2]$ sec. The step response of the original model and the ROMs is computed using MATLAB's \textit{``step''} command. The absolute error, i.e., $||y-\hat{y}_r||$, is plotted in Fig. \ref{fig:1}.
It can be seen that TLPORK compares well with TLBT and TLIRKA. Note that unlike TLBT and TLIRKA, the stability of the ROM is guaranteed in TLPORK and O-TLPORK because the poles of the ROM are selected by the user, (i.e., as the mirror images of the interpolation points,) and the interpolation points must lie in the right-half of the $s$-plane. The $\mathcal{H}_{2,t}$ and $\mathcal{H}_\infty$ norms of the error are tabulated in Table \ref{tab2}.
\begin{table}[!h]
	\centering
	\caption{$\mathcal{H}_{2,t}$-norm Error}\label{tab2}
	\begin{tabular}{|c|c|c|}
		\hline
		Technique & $||H(s)-\hat{H}_r(s)||_{\mathcal{H}_{2,t}}$&$||H(s)-\hat{H}_r(s)||_{\mathcal{H}_{\infty}}$ \\ \hline
		BT   & $7.6520\times 10^{-6}$&$3.6950\times 10^{-6}$\\
		TLBT&$1.1589\times 10^{-6}$&$0.0742$\\
        A-TLBT&$1.1545\times 10^{-6}$&$0.0673$\\
		IRKA&$7.8103\times 10^{-6}$&$3.3822\times 10^{-6}$\\
        ISRKA&$7.7093\times 10^{-6}$&$3.4314\times 10^{-6}$\\
		PORK&$7.8103\times 10^{-6}$&$3.3822\times 10^{-6}$\\
		TLIRKA&$1.1545\times 10^{-6}$&$0.0673$\\
        TLPORK&$1.1347\times 10^{-6}$&$0.0673$\\
        O-TLPORK&$1.1347\times 10^{-6}$&$0.0673$\\\hline
	\end{tabular}
\end{table}
Interestingly, TLPORK and O-TLPORK ensure less $\mathcal{H}_{2,t}$-norm error than TLIRKA in this example. This is not a surprising result as TLIRKA does not yield an optimal ROM but only tends to yield an optimal ROM and nearly satisfies the optimality conditions. The $\mathcal{H}_\infty$-norm error shows the accuracy of the ROM over the entire (frequency and) time horizon. One can notice that the infinite horizon MOR techniques outclass the time-limited MOR techniques when $\mathcal{H}_\infty$-norm errors are considered, which is intuitive.

\begin{figure}[!h]
\centering
\includegraphics[width=8.2cm]{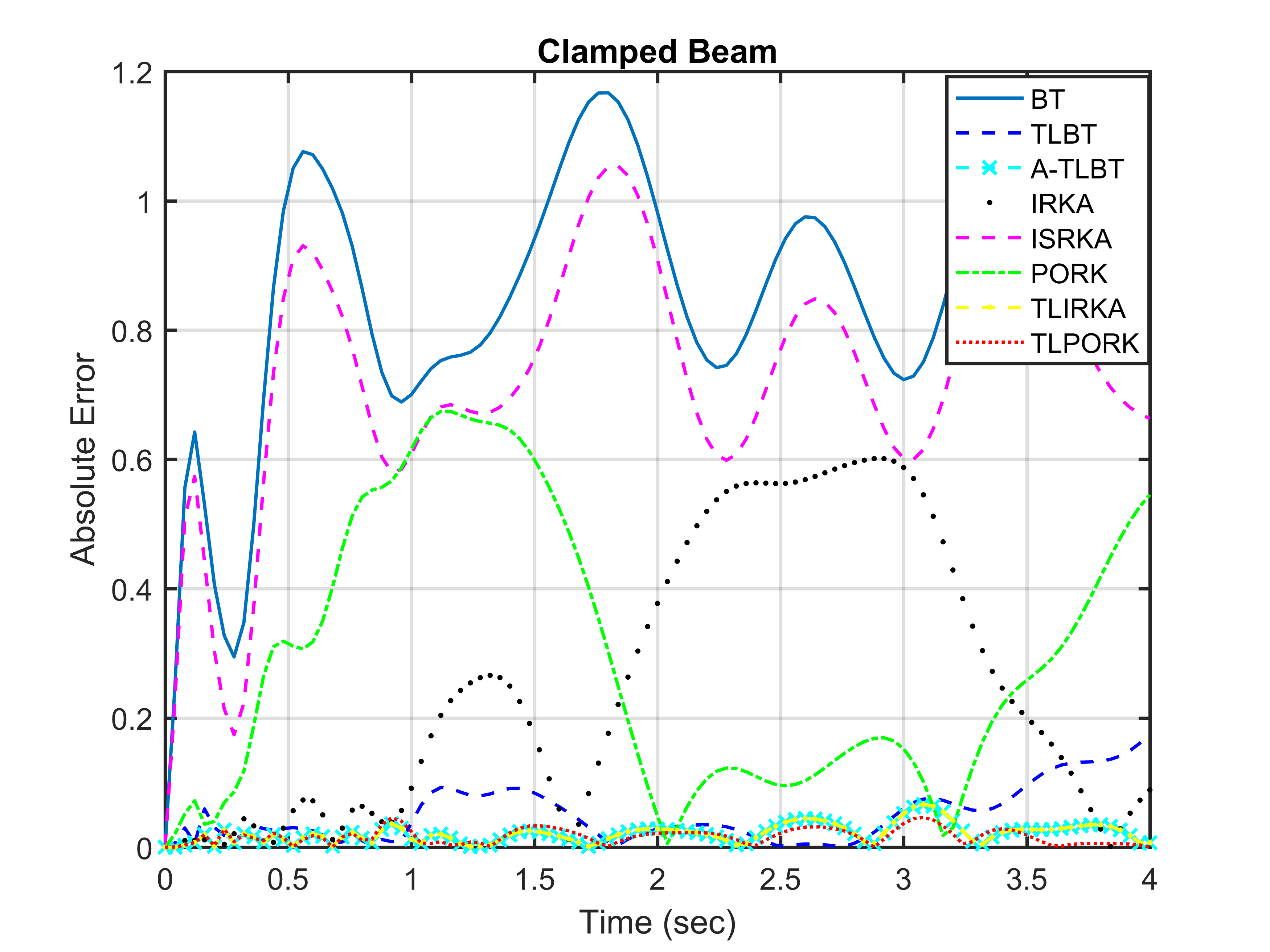}
\caption{Absolute error in the step response}\label{fig:2}
\end{figure}

\textbf{Clamped beam:} This is a $348^{th}$ order SISO model taken from \cite{chahlaoui2002collection}. In this experiment, a unit step input is applied to the system at $0$ sec for four seconds. The goal of the experiment is to construct a ROM that accurately mimics the original system during these four seconds when the same input is applied. A $12^{th}$ order ROM is obtained using BT, TLBT, IRKA, ISRKA, PORK, TLIRKA, TLPORK, O-TLPORK, and A-TLBT. The desired time interval is specified as $[0,4]$ sec. The step response of the original model and the ROMs is computed using MATLAB's \textit{``step''} command. The absolute error, i.e., $||y-\hat{y}_r||$, is plotted in Fig. \ref{fig:2}.
It can be seen that TLPORK compares well with TLBT and TLIRKA. The $\mathcal{H}_{2,t}$ and $\mathcal{H}_\infty$ norms of the error are tabulated in Table \ref{tab3}.
\begin{table}[!h]
	\centering
	\caption{$\mathcal{H}_{2,t}$-norm Error}\label{tab3}
	\begin{tabular}{|c|c|c|}
		\hline
		Technique & $||H(s)-\hat{H}_r(s)||_{\mathcal{H}_{2,t}}$& $||H(s)-\hat{H}_r(s)||_{\mathcal{H}_{\infty}}$ \\ \hline
		BT   & $3.3993$&$2.3748$\\
		TLBT&$1.0228$&$4.4856\times 10^3$\\
        A-TLBT&$0.5455$&$4.5769\times 10^3$\\
		IRKA&$1.3301$&$19.8369$\\
        ISRKA&$3.3260$&$2.2517$\\
		PORK&$1.3785$&$20.0268$\\
		TLIRKA&$0.5455$&$4.5769\times 10^3$\\
        TLPORK&$0.5191$&$4.5763\times 10^3$\\
        O-TLPORK&$0.5191$&$4.5763\times 10^3$\\\hline
	\end{tabular}
\end{table}
It can be seen that TLPORK and O-TLPORK ensure the least $\mathcal{H}_{2,t}$-norm error. Again, as expected, the infinite horizon MOR techniques outclass the time-limited MOR techniques when $\mathcal{H}_\infty$-norm errors are considered.

\textbf{Brazil Interconnected Power System:} This is a $3077^{th}$ order MIMO model taken from \cite{rommes2009computing} with four inputs and four outputs. A unit step input is applied to the system at $0$ sec for three seconds. Again, the goal of the experiment is to construct a ROM that accurately mimics the original system during these three seconds when the same input is applied. A $25^{th}$ order ROM is obtained using BT, TLBT, IRKA, ISRKA, PORK, TLIRKA, TLPORK, O-TLPORK, and A-TLBT. The desired time interval is specified as $[0,3]$ sec. We avoid plotting the step response in this case as it is a MIMO system with $16$ response plots. The $\mathcal{H}_{2,t}$ and $\mathcal{H}_\infty$ norms of the error are tabulated in Table \ref{tab4}. It can be seen that TLPORK and O-TLPORK ensure the least $\mathcal{H}_{2,t}$-norm error.
\begin{table}[!h]
	\centering
	\caption{$\mathcal{H}_{2,t}$-norm Error}\label{tab4}
	\begin{tabular}{|c|c|c|}
		\hline
		Technique & $||H(s)-\hat{H}_r(s)||_{\mathcal{H}_{2,t}}$&$||H(s)-\hat{H}_r(s)||_{\mathcal{H}_{\infty}}$ \\ \hline
		BT   & $24.2988$&$4.0541$\\
		TLBT& $1.4003$&$9.1142$\\
        A-TLBT&$0.7537$&$2.9662$\\
		IRKA&$1.3156$&$2.1406$\\
        ISRKA&$1.0945$&$2.3048$\\
		PORK&$1.3156$&$2.1406$\\
		TLIRKA&$0.7585$&$2.8972$\\
        TLPORK&$0.7478$&$3.0644$\\
        O-TLPORK&$0.7421$&$3.0745$\\\hline
	\end{tabular}
\end{table}
\section{Conclusion}
We present an iteration-free tangential interpolation algorithm, which enforces a subset of the first-order optimality conditions for time-limited $\mathcal{H}_2$-MOR problem. The proposed algorithm can also preserve the desired modes and their associated residues in the ROM like modal truncation. The Krylov subspace-based implementation ensures its applicability to large-scale systems. We also present an adaptive framework of the proposed algorithm, which increases the order the ROM until the desired tolerance in the error is achieved, and it ensures that the error decreases monotonically after each step. The numerical results confirm the theory presented in the paper.
\section{Acknowledgment}
This work is supported by National Natural Science Foundation of China under Grant (No. 61873336, 61873335), and supported in part by 111 Project (No. D18003).

\end{document}